\documentstyle[a4wide,fleqn,epsf]{elsart}

\newcommand{\bra}[1]{\left\langle #1 \right|}
\newcommand{\ket}[1]{\left| #1 \right\rangle}

\begin{document}

\begin{frontmatter}

\title{
       Effective interactions for the nuclear shell model
      }

\author[cph,oslo]{M.\ Hjorth-Jensen}

\address[cph]{NORDITA, Blegdamsvej 17, DK-2100 K\o benhavn \O, Denmark}

\address[oslo]{Department of Physics, University of Oslo, N-0316 Oslo, Norway}

\maketitle

\begin{abstract}
Various perturbative and non-perturbative many-body techniques
are discussed in this work. Especially, we will focus on the summation
of so-called Parquet diagrams with emphasis on applications to finite nuclei.
Here, the subset of two-body Parquet equations will be discussed. A 
practical implementation of the corresponding equations
for studies of effective interactions
for finite nuclei is outlined. 
\end{abstract}


\end{frontmatter}
\newpage

\tableofcontents

\section{Introduction}

Traditional shell-model studies 
have recently received a renewed
interest through large scale shell model calculations
in both the $1p0f$ shell and the $2s1d0g_{7/2}$ shells with
the inclusion of the $0h_{11/2}$ intruder state as well. 
It is now therefore fully possible to perform large-scale 
shell-model investigations
and study the excitation spectra 
for systems with 
some 10 million basis states. With recent advances in
Monte Carlo methods one is also 
able to enlarge the dimensionality
of the systems under study considerably,
and important information on e.g., ground state properties
has thereby been obtained.

An important feature of such large scale calculations
is that it allows one to probe the underlying many-body
physics in a hitherto unprecedented way.
The crucial starting point in all such shell-model 
calculations is
the derivation of an effective interaction, be it
either an approach based on a microscopic theory
starting from the free nucleon-nucleon ($NN$) interaction or a more 
phenomenologically determined interaction. 
In shell-model studies of e.g., the Sn isotopes, one may have
up to 31 valence particles or holes interacting via e.g.,
an effective two-body interaction. The results of such 
calculations can therefore yield, when compared with 
the availiable body of experimental data, critical
inputs to the underlying theory of the effective interaction.
Until very recently, realistic shell-model effective interactions have
mainly been applied to nuclei with two or a few valence particles beyond
closed shells, such as the oxygen and calcium isotopes. Thus, by going to the
tin isotopes, in which the major neutron shell between neutron numbers 50 and
82 is being filled beyond the $^{100}$Sn closed shell core, we have the opportunity
of testing the potential of large-scale 
shell-model calculations as well as the reliability of
realistic effective interactions in systems with many valence particles. 

Clearly, although the $NN$  interaction is of short
but finite range, with typical interparticle
distances of the order of $1\sim 2$ fm, there are  
indications from both studies of few-body systems such as the triton and
infinite nuclear matter, that at least three-body
interactions, both real and effective ones, may be of
importance. 
Thus, with many valence nucleons present, such
large-scale shell-model calculations may
tell us how well e.g., an effective interaction
which only includes two-body terms does in
reproducing properties such as excitation spectra and
binding energies. 

This work deals therefore with various ways of 
deriving the effective
interaction or effective operator needed
in shell-model calculations, starting from the
free $NN$  interaction.
Normally, the problem of deriving such effective operators and interactions are solved
in a limited space, the so-called model space, which is a subspace of the
full Hilbert space. The effective operator and interaction
theory is then introduced in order to
systematically take into account contributions from the complement
(the excluded space) of the chosen model space. Several formulations
for such expansions of effective operators and interactions exit in the literature, following
time-dependent or time-independent
perturbation theory \cite{so95,brandow67,ko90,hko95,lm85,so84}.  
Formulations like the coupled-cluster
method or exponential ansatz 
\cite{lm85,arponen97,lk72a,lk72b,zabolitzky74,klz78,ticcm,hm98},
the summation of the Parquet class of diagrams 
\cite{dm64,nozieres,babu,jls82,br86,scalapino,ym96,dya97}, or
the so-called $\hat{Q}$-box method with the folded-diagram
formulation of Kuo and co-workers \cite{ko90,hko95} have been extensively
applied to systems in nuclear, atomic, molecular and solid-state physics.
Here we will focus on the above-mentioned $\hat{Q}$-box approach
and the summation of the so-called Parquet diagrams. 
For the description of other many-body methods such as the Hypernetted-chain or the 
correlated basis function \cite{adelchi98} methods, monte-carlo related 
methods
\cite{david97,vijay}, Unitary-correlation operator method \cite{so84} etc.\ 
see other contributions in this volume.

The
$\hat{Q}$-box has been introduced in Rayleigh-Schr\"odinger perturbation
theory as the definition of all non-folded diagrams to a given order in
the expansion parameter, in nuclear physics the so-called $G$-matrix.
The $G$-matrix renders the free $NN$  interaction $V$, which
is repulsive at small internucleon distances,  tractable
for a perturbative analysis through the summation of ladders diagrams
to infinite order. Stated differently, the $G$-matrix, through the
solution of the Bethe-Brueckner-Goldstone equation, accounts for the
short-range correlations involving high-lying states.
Folded diagrams are a class of diagrams which arise
due to the removal of the dependence of the exact model-space energy
in the Brillouin-Wigner perturbation expansion. Through the $\hat{Q}$-box
formulation and its derivatives, this set of diagrams can easily be summed
up. 

In addition to the evaluation of folded diagrams and the inclusion
of ladder diagrams to infinite order included in the $G$-matrix, there
are other classes of diagrams which can be summed up.
These take into account
the effect of long-range correlations involving low-energy excitations.
A frequently applied formalism is
the Tamm-Dancoff (TDA) or the random-phase
(RPA) approximations. In their traditional formulation one allows
for the summation of all particle-hole excitations, both forward-going
and backward going insertions.
This set of diagrams, as formulated by Kirson \cite{kirson74} and reviewed
in Ref.\ \cite{eo77}, should account for correlations  arising
from collective particle-hole correlations. Another possibility,
is to include any number of particle-particle and hole-hole 
correlations in the screening of particle-hole correlations. 
The inclusion of these kind of correlations is conventionally labelled
particle-particle (pp) RPA. It has been used both in nuclear matter
studies \cite{angels88,rpd89,yhk86,syk87}
and in evaluations of ground state properties of closed-shell
nuclei \cite{hmtk87,emm91,hmm95}.
Recently, Ellis, Mavromatis and M\"uther \cite{emm91,hmm95} have
extended the pp RPA to include the particle-hole (ph) RPA, though
screening of two-particle-one-hole (2p1h) and two-hole-one-particle
(2h1p) vertices was not included.
The latter works can be viewed as a step towards the full summation of the
Parquet class of diagrams. 
The summation of the Parquet diagrams entails a self-consistent
summation of both particle-particle and hole-hole ladder diagrams
and particle-hole diagrams. Practical solutions to this many-body
scheme for finite nuclei will be discussed here.

This work falls in six sections.
In the next section we present various definitions pertinent
to the determination of effective interactions, with an emphasis
on perturbative methods.
The resummation of the ladder type of 
diagrams is then presented in section \ref{sec:sec3}.
In that section we also discuss the summation of so-called
folded diagrams which arise in the evaluation of 
valence space effective interactions. Further perturbative 
corrections are also discussed and selected results
for light nuclei in the $1s0d$ and $1p0f$ shells and for
Sn isotopes are presented.

The summation of the TDA and RPA class of diagrams is discussed in
section \ref{sec:sec4}. Other screening corrections are also
discussed in that section. The self-consistent approach to the 
summation of both ladder type diagrams and screening terms 
through the solution    
of equations for the Parquet class of diagrams will be presented
in section \ref{sec:sec5}. 
Concluding remarks are given in section \ref{sec:sec6}.

\section{Perturbative methods}
\label{sec:sec2}

In order to derive a microscopic approach to the effective interaction and/or operator 
within the framework of perturbation theory, we need to introduce various
notations and definitions pertinent to the methods exposed.
In this section we review how to calculate an effective 
operator within the framework of 
degenerate Rayleigh-Schr\"{o}dinger
(RS) perturbation theory \cite{ko90,lm85}. 

It is common practice in perturbation theory to reduce the infinitely
many degrees of freedom of the Hilbert space to those represented
by a physically motivated subspace, the model space.
In such truncations of the Hilbert space, the notions of a projection
operator $P$ onto the model space and its complement $Q$ are
introduced. The projection operators defining the model and excluded
spaces are defined by
\begin{equation}
        P=\sum_{i=1}^{D} \left|\Phi_i\right\rangle
        \left\langle\Phi_i\right |,
\end{equation}
and
\begin{equation}
        Q=\sum_{i=D+1}^{\infty} \left|\Phi_i\right\rangle
        \left\langle\Phi_i\right |,
\end{equation}
with $D$ being the dimension of the model space, and $PQ=0$, $P^2 =p$,
$Q^2 =Q$ and $P+Q=I$. The wave functions $\left|\Phi_i\right\rangle$ 
are eigenfunctions
of the unperturbed hamiltonian $H_0 = T+U$, where $T$ is the kinetic
energy and $U$ and appropriately chosen one-body potential, that of the
harmonic oscillator (h.o.) in most calculations. The full hamiltonian
is then rewritten as $H=H_0 +H_1$ with $H_1=V-U$, $V$ being e.g.\ the
$NN$    interaction. The eigenvalues
and eigenfunctions of the full hamiltonian are denoted by
$\left|\Psi_{\alpha}\right\rangle$
and $E_{\alpha}$,
\begin{equation}
                H\left|\Psi_{\alpha}\right\rangle= 
                E_{\alpha}\left|\Psi_{\alpha}\right\rangle.
\end{equation}
Rather than solving the full Schr\"{o}dinger equation above, we define
an effective hamiltonian acting within the model space such
that
\begin{equation}
               PH_{\mathrm{eff}}P\left|\Psi_{\alpha}\right\rangle=
               E_{\alpha}P\left|\Psi_{\alpha}\right\rangle=
              E_{\alpha}\left|\Phi_{\alpha}\right\rangle
\end{equation}
where $\left|\Phi_{\alpha}\right\rangle=P\left|\Psi_{\alpha}\right\rangle$
is the projection of the full wave function
onto the model space, the model space wave function.
In RS perturbation theory, the effective interaction
$H_{\mathrm{eff}}$ can be written out order by order in the 
interaction $H_1$ as
\begin{equation}
               PH_{\mathrm{eff}}P=PH_1P +PH_1\frac{Q}{e}H_1 P+
               PH_1\frac{Q}{e}H_1 \frac{Q}{e}H_1 P+\dots,
               \label{eq:effint}
\end{equation}
where terms of third and higher order also
include the aforementioned folded diagrams. 
Further, $e=\omega -H_0$,
where $\omega$ is the so-called starting energy, defined as the unperturbed
energy of the interacting particles..
Similarly,
the exact wave
function $\left|\Psi_{\alpha}\right\rangle$
can now be written in terms of the model space wave function as
\begin{equation}
                \left|\Psi_{\alpha}\right\rangle=
                \left|\Phi_{\alpha}\right\rangle+
                \frac{Q}{e}H_1\left|\Phi_{\alpha}\right\rangle
                +\frac{Q}{e}H_1\frac{Q}{e}H_1\left|\Phi_{\alpha}\right\rangle
                +\dots
                \label{eq:wavef}
\end{equation}
The wave operator is often expressed as
\begin{equation}
              \Omega = 1 +\chi,
\end{equation}
where $\chi$ is known as the correlation operator. The correlation
operator generates the component of the wave function in the $Q$-space
and must therefore contain at least one perturbation. Observing
that $P\Omega P = P$, we see that the correlation operator $\chi$
has the properties
\begin{equation}
               P\chi P = 0, \hspace{1cm} Q\Omega P = 
              Q\chi P =\chi P. \label{eq:chi1}
\end{equation}
Since  $\left|\Psi_i\right\rangle=\Omega\left|\Psi_i^{M}\right\rangle$ 
determines the wave operator
only when it operates to the right on the model space, i.e., only the
$\Omega P$  part is defined, the term $\Omega Q$
never appears in the theory,
and we could therefore add the conditions $Q\chi Q =0$ and $P\chi Q =0$
to Eq.\ (\ref{eq:chi1}). This leads to the following choice for $\chi$
\begin{equation}
                   \chi = Q\chi P. \label{eq:chi2}
\end{equation}
This has been the traditional choice in perturbation theory \cite{so95,lm85}.

\begin{figure}[hbtp]
      \setlength{\unitlength}{1mm}
      \begin{picture}(100,60)
      \put(35,0){\epsfxsize=8cm \epsfbox{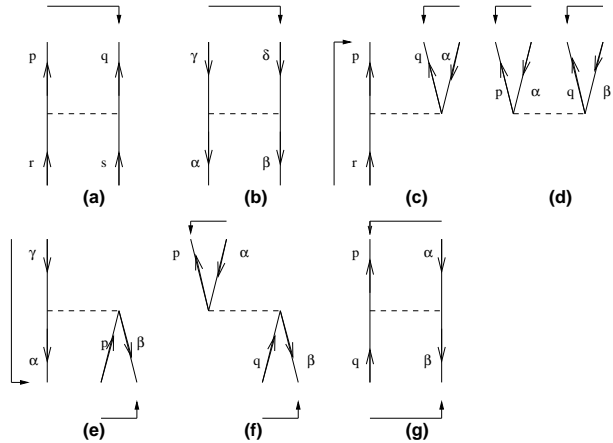}}
      \end{picture}
      \caption{The various vertices to first order in the interaction
               $V$ which contribute to the wave operator
               $\Omega =1+\chi$. Hartree-Fock
               terms are not included. Possible hermitian conjugate 
                diagrams are also not shown. Indicated are also possible
               angular momentun coupling orders.}
      \label{fig:wavef1}
\end{figure}

The wave operator $\Omega$ can then be ordered in terms of the number
of interactions with the perturbation $H_1$
\begin{equation}
              \Omega = 1 +\Omega^{(1)} + \Omega^{(2)}+\dots ,
\end{equation}
where $\Omega^{(n)}$ means that we have $n$ $H_1$ terms. 
Explicitely, the above
equation reads
\begin{eqnarray}
         \Omega\left|\Phi_i\right\rangle=
         &{\displaystyle\left|\Phi_i\right\rangle
         +\sum_{\alpha}\frac{\left|\alpha\right\rangle
         \left\langle\alpha\right|
          V\left|\Phi_i\right\rangle}{\varepsilon_i -\varepsilon_{\alpha}}
         +\sum_{\alpha\beta}\frac{\left|\alpha\right\rangle
        \left\langle\alpha\right| V
         \left|\beta\right\rangle\left\langle\beta\right| V
         \left|\Phi_i\right\rangle }
         {(\varepsilon_i-\varepsilon_{\alpha})
       (\varepsilon_i-\varepsilon_{\beta})} }\\   \label{eq:wavefu}\nonumber
&       {\displaystyle  -\sum_{\alpha j}\frac{\left|\alpha\right\rangle
       \left\langle\alpha\right|
         V\left|\Phi_j\right\rangle
        \left\langle\Phi_j\right| V\left|\Phi_i\right\rangle}
       {(\varepsilon_i-\varepsilon_{\alpha})
      (\varepsilon_i-\varepsilon_{j})} }
       +\dots ,
\end{eqnarray}
where $\varepsilon$ are the unperturbed energies of the $P$-space
and $Q$-space states defined by $H_0$.
The greek letters refer to
$Q$-space states, whereas a latin letter refers to model-space
states.   The second term
in the above equation corresponds to $\Omega^{(1)}$ while the third
and fourth define $\Omega^{(2)}$.
Note that the fourth term diverges
in case we have a degenerate or nearly degenerate model space. It is
actually divergencies like these which are to be removed by the folded
diagram procedure to be discussed in the next section. Terms like these
arise due to the introduction of an energy independent perturbative
expansion. Conventionally, the various contributions to the
perturbative expansion are represented by Feynman-Goldstone diagrams.
In Fig.\ \ref{fig:wavef1} we display the topologically distinct
contributions to first order in the
interaction $V$ to
the wave operator Eq.\ (\ref{eq:wavefu}). These diagrams all
define the correlation operator $\chi$ to first order in $V$.
Diagrams with Hartree-Fock contributions
and single-particle potential terms 
are not included. The possible 
renormalizations of these diagrams will be discussed
in the next three sections.
The reader should note that with respect to
the nomenclature in Eq.\ (\ref{eq:wavefu}), we will hereafter 
employ the following notation in our discussion of various diagrams
and vertex renormalizations:
\begin{itemize}
\item roman letters $p,q,r,s,t,\dots$ refer to particle single-particle
states, either
within the model-space or from the excluded space. An arrow pointing
upwards represents such a particle state. 
\item greek letters $\alpha ,\beta ,\gamma ,\delta \dots$ refer to hole
single-particle states. An arrow pointing downward is a hole state.
\end{itemize}

\subsection{Expressions for the wave operator}

\begin{figure}[hbtp]
      \setlength{\unitlength}{1mm}
      \begin{picture}(100,40)
      \put(35,0){\epsfxsize=7cm \epsfbox{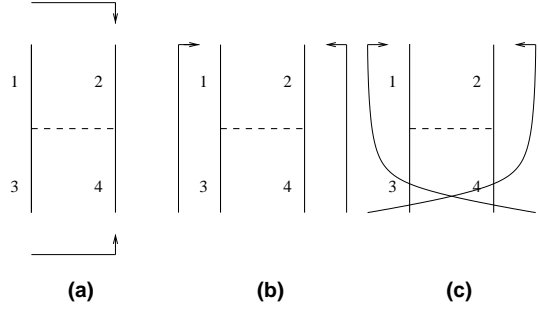}}
      \end{picture}
      \caption{Coupling order for the $[12]$ (a), $[13]$ (b) and
               $[14]$ (c) channels.}
      \label{fig:channelsdef}
\end{figure}

We end this section with the equations for the diagrams
in Fig.\ \ref{fig:wavef1} representing $\chi$ to first order in $V$.
Moreover, in order to introduce the various channels needed to sum the 
Parquet class of diagrams, we will find it convenient here to
classify these channels in terms of angular momentum
recouplings. Later on, we will also introduce the pertinent
definitions of energy and momentum variables in the various channels.
The nomenclature we will follow in our labelling is that
of Blaizot and Ripka, see Ref.\ \cite{br86} chapter 15.
All matrix elements in the definitions below 
are antisymmetrized and unnormalized.
The first channel is the $[12]$ channel, or the $s$-channel
in field theory, and its angular momentum coupling order
is depicted in Fig.\ \ref{fig:channelsdef}. 
In this figure 
we do not distinguish between particles and holes, all single-particle
labels $1,2,3,4$ can represent either a hole or particle
single-particle state. It is the coupling order which is
of interest here.  
The matrix element $V^{[12]}$ in the $[12]$ channel is then 
\begin{equation}
       V_{1234J}^{[12]}
       =\left\langle (12)J\right | V
       \left | (34)J\right\rangle,
       \label{eq:12channel}
\end{equation} 
meaning that the single-particle state $1(3)$ couples to the state
$2(4)$ to yield a total angular momentum $J$. 
This channel is commonly denoted as the particle-particle (pp)
or particle-particle-hole-hole (pphh) channel, meaning that when
we will sum classes of diagrams to infinite order in this channel, the only
intermediate states which are allowed are those of a pphh character,
coupled to a final $J$ in the above order.
In the next section we will explicitely discuss ways to evaluate
the equations for the $[12]$ channel. 
This coupling order is also the standard way of representing
effective interactions for shell-model
calculations.
If we now specialize to particles and holes (these matrix 
elements were shown in Fig.\ \ref{fig:wavef1}) we obtain for the case  
with particle states only, diagram (a),
\begin{equation}
      V_{\mathrm{2p}}=V_{pqrs J}^{[12]}=
       \left\langle (pq)J\right | V\left | (rs)J\right\rangle.
       \label{eq:2pv}
\end{equation}
The corresponding one for holes only, diagran (b), is
\begin{equation}
      V_{\mathrm{2h}}=V_{\alpha\beta\gamma\delta J}^{[12]}=
       \left\langle (\alpha\beta)J\right | V
       \left | (\gamma\delta)J\right\rangle.
       \label{eq:2hv}
\end{equation}
Thus, in the forthcoming discussion, we will always employ as our 
basic notation for a matrix element that of the $[12]$ channel,
meaning that matrix elements of the other two channels
can always be rewritten in terms of those in $[12]$ channel
We see this immediately by looking at the expression for the 
matrix element in the $[13]$ channel, the $t$-channel in field
theory, see Fig.\ 
\ref{fig:channelsdef}(b). Here the single-particle state $3(4)$
couples to the single-particle state $1(2)$\footnote{In a Goldstone-Feynman
diagram in an angular momentum representation, the coupling direction
will always be from incoming single-particle states to outgoing
single-particle states.}.
Through simple angular momentum algebra we have 
\begin{equation}
      V_{1234J}^{[13]}=
      {\displaystyle \sum_{J'}}(-)^{j_1+j_4+J+J'}\hat{J'}^2
      \left\{
      \begin{array}{ccc}
       j_3&j_1&J\\j_2&j_4&J'
      \end{array}
       \right\}V_{1234J}^{[12]},
       \label{eq:13channel}
\end{equation}
where the symbol with curly brackets represents a $6j$-symbol and
$\hat{J'}=\sqrt{2J'+1}$.
In a similar way we can also express the matrix
element in the $[14]$ channel, the $u$-channel in field theory,
through  
\begin{equation}
       V_{1234J}^{[14]}=
      {\displaystyle \sum_{J'}}(-)^{j_1+j_4+J+2j_3}\hat{J'}^2
      \left\{
      \begin{array}{ccc}
       j_4&j_1&J\\j_2&j_3&J'
      \end{array}
       \right\}
       V_{1234J}^{[12]}.
       \label{eq:14channel}
\end{equation}
It is also possible to have the inverse relations or to express
e.g., the $[14]$ channel through the $[13]$ channel as
\begin{equation}
       V_{1234J}^{[14]}=
      {\displaystyle \sum_{J'}}(-)^{2j_1+2j_2+2j_3}\hat{J'}^2
      \left\{
      \begin{array}{ccc}
       j_4&j_1&J\\j_3&j_2&J'
      \end{array}
       \right\}
       V_{1234J}^{[13]}.
       \label{eq:1413channel}
\end{equation}
The matrix elements defined in Eqs.\ 
(\ref{eq:12channel})-(\ref{eq:1413channel}) and the inverse relations 
are the 
starting points for various resummation of diagrams.
In the next section we will detail ways of solving equations
in the $[12]$ channel, whereas various approximations for the 
$[13]$ channel and $[14]$ channel such as the TDA and RPA 
and  vertex and propagator renormalization schemes 
will be discussed in section \ref{sec:sec4}. Finally, how to
merge self-consistently  all three channels will be discussed
in section \ref{sec:sec5}.

We end this section by giving the expressions in an angular momentum
basis\footnote{Note that we only include angular momentum factors,
other 
factors coming from the diagram rules\cite{kstop81}, 
like number of hole lines,
number of closed loops etc.\ are omitted here.} for the remaining 
diagrams of Fig.\ \ref{fig:wavef1}. The coupling order is indicated
in the same figure. 

Thus, the 2p1h vertex $V_{\mathrm{2p1h}}$, 
diagram (c) in  Fig.\  \ref{fig:wavef1}, 
is coupled
following the prescription of the $[13]$ channel and reads
\begin{equation}
      V_{\mathrm{2p1h}}=V_{pqr\alpha J}^{[13]}=
      {\displaystyle \sum_{J'}}(-)^{j_{\alpha}+j_p+J+J'}\hat{J'}^2
      \left\{
      \begin{array}{ccc}
       j_r&j_p&J\\j_q&j_{\alpha}&J'
      \end{array}
       \right\}
       V_{pqr\alpha J'}^{[12]}.
       \label{eq:2p1hv}
\end{equation}
The 2p2h ground-state correlation $V_{\mathrm{2p2h}}$, diagram (d),
 which will enter
in the RPA summation discussed in section \ref{sec:sec4} is given by,
the coupling order is that of the $[13]$ channel, 
\begin{equation}
      V_{\mathrm{2p2h}}=V_{pq\alpha\beta J}^{[13]}=
      {\displaystyle \sum_{J'}}(-)^{j_{\beta}+j_p+J+J'}\hat{J'}^2
      \left\{
      \begin{array}{ccc}
       j_{\alpha}&j_p&J\\j_q&j_{\beta}&J'
      \end{array}
       \right\}
       V_{pq\alpha\beta J'}^{[12]}.
       \label{eq:2p2hv}
\end{equation}
The 2h1p vertex $V_{\mathrm{2h1p}}$, diagram (e), 
still in the representation of 
the $[13]$ channel, is defined as
\begin{equation}
      V_{\mathrm{2h1p}}=V_{\alpha\beta\gamma p J}^{[13]}=
      {\displaystyle \sum_{J'}}(-)^{j_{\alpha}+j_p+J+J'}\hat{J'}^2
      \left\{
      \begin{array}{ccc}
       j_{\gamma}&j_{\alpha}&J\\j_{\beta}&j_p&J'
      \end{array}
       \right\}
       V_{\alpha\beta\gamma p J'}^{[12]}.
       \label{eq:2h1pv}
\end{equation}
Note well that the vertices of Eqs.\ (\ref{eq:2p1hv})-(\ref{eq:2h1pv})
and their respective 
hermitian conjugates can all be expressed in the $[14]$ channel
or $[12]$ channel as well.
However, it is important to note that the expressions in the various
channels are different, and when solving the equations for the various 
channels, the renormalizations will be different. As an example, consider
the two particle-hole vertices $V_{\mathrm{ph}}$ 
of Fig.\ \ref{fig:wavef1}, i.e., diagrams (f) and (g).
Diagram (g) is just the exchange diagram of (f) when seen in the 
$[12]$ channel. However, if (f) is coupled as in the $[13]$ channel,
recoupling this diagram to the $[14]$ channel will not give 
two particle-hole
two-body states coupled to a final $J$ but rather 
a particle-particle two-body state and a hole-hole  two-body state.
But why bother at all about such petty details? The problem arises when we 
are to  sum diagrams in the $[13]$ channel and $[14]$ channel.  
In the $[12]$ channel we allow only particle-particle and hole-hole
intermediate states, whereas in the $[13]$ channel and $[14]$ channel
we allow only particle-hole intermediate states, else we may risk
to double-count various contributions. 
If we therefore recouple
diagram (f) to the $[14]$ representation, this contribution
does not yield an intermediate particle-hole state
in the $[14]$ channel.
Thus, diagram (f), whose expression is 
\begin{equation}
      V_{\mathrm{ph}}=V_{p\beta \alpha q J}^{[13]}=
      {\displaystyle \sum_{J'}}(-)^{j_p+j_q+J+J'}\hat{J'}^2
      \left\{
      \begin{array}{ccc}
       j_{\alpha}&j_p&J\\j_{\beta}&j_q&J'
      \end{array}
       \right\}
       V_{p\beta \alpha q J'}^{[12]},
       \label{eq:ph13}
\end{equation}
yields a particle-hole contribution only in the $[13]$ channel,
whereas the exchange diagram (g), which reads
\begin{equation}
      V_{\mathrm{ph}}=V_{p\beta q\alpha J}^{[14]}=
      {\displaystyle \sum_{J'}}(-)^{2j_q+j_{\alpha}+j_p+J}\hat{J'}^2
      \left\{
      \begin{array}{ccc}
       j_{\alpha}&j_p&J\\j_{\beta}&j_{q}&J'
      \end{array}
       \right\}
       V_{p\beta q\alpha J'}^{[12]},
       \label{eq:ph14}
\end{equation}
results in the corresponding particle-hole contribution in the 
$[14]$ channel.
In electron gas theory, the latter expression 
is often identified as the starting point for the self-screening
of the exchange term. In the discussion of the TDA series in 
section \ref{sec:sec4} we will give the expressions for the screening
corrections based on Eqs.\ (\ref{eq:ph13}) and (\ref{eq:ph14}). 

An important aspect to notice in connection with the latter
equations and the discussions in  sections \ref{sec:sec4}
and \ref{sec:sec5} is that 
\begin{equation} 
    V_{p\beta q\alpha J}^{[14]}=-V_{p\beta \alpha q J}^{[13]},
\end{equation}
i.e., just the exchange diagram, as it should be.
This is however important to keep in mind, since we later
on will sum explicitely sets of diagrams in the
$[13]$ channel and the $[14]$ channel, implying thereby that
we will obtain screening and vertex corrections
for  both direct and exchange  diagrams. 

\section{Summation of diagrams in the $[12]$ channel}
\label{sec:sec3}

\begin{figure}[hbtp]
      \setlength{\unitlength}{1mm}
      \begin{picture}(100,60)
      \put(35,0){\epsfxsize=7cm \epsfbox{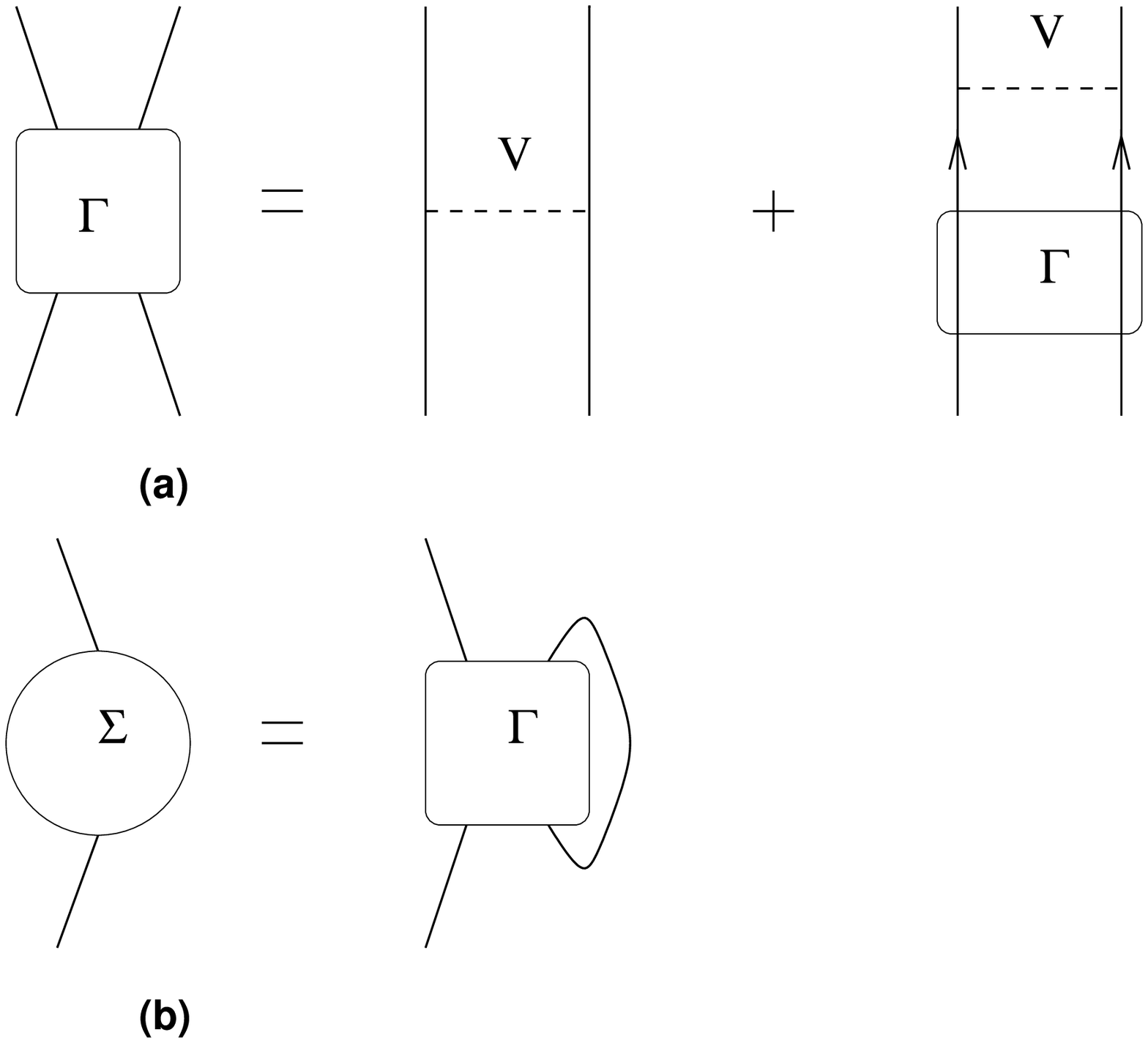}}
      \end{picture}
      \caption{(a) represents the two-body vertex $\Gamma$ function while (b)  
               represents the self-energy $\Sigma$.}
      \label{fig:selfcons12}
\end{figure}

In order to write down the equation for the renormalized 
interaction $\Gamma^{[12]}$ in the 
$[12]$ channel we need first to present some further definitions.
We will also assume that the reader has some familiarity with the theory 
of Green's function. Thorough discussions of such topics
can be found in the recent reviews of Dickoff and M\"uther \cite{dm92}
and Kuo and Tzeng \cite{kt94}. In our presentation below we will
borrow from these works and the monograph of Blaizot and Ripka \cite{br86}.
The vertex $\Gamma^{[12]}$ is in lowest order identical with the 
interaction $V^{[12]}$ and obeys also the same symmetry relations
as $V$, i.e.,
\begin{equation}
     \Gamma^{[12]}_{1234J}=\Gamma^{[12]}_{2143J}=-\Gamma^{[12]}_{2134J}=
     \Gamma^{[12]}_{1243J}.
     \label{eq:symproperties}
\end{equation}
We also need to define energy variables. Since we are going to 
replace the interaction $V$ with the $G$-matrix, or certain
approximations to it,  defined below in all
of our practical calculations, the momentum variables are already
accounted for in $G$. The basis will be that of harmonic oscillator 
wave functions, and the labels $1234$ will hence refer to oscillator
quantum numbers, which in turn can be related to the momentum
variables. The labels $1234$, in addition to representing 
single-particle quantum numbers, define also the energy of the single-particle
states. With a harmonic oscillator basis, the starting point for the 
single-particle energies $\varepsilon_{1,2,3,4}$ are the unperturbed 
oscillator energies. When iterating the equations for $\Gamma^{[12]}$,
self-consistent single-particle energies can be introduced. 
The total energy in the $[12]$ channel $s$ is
\begin{equation}
    s=\varepsilon_1+\varepsilon_2=\varepsilon_3+\varepsilon_4.
    \label{eq:energy12}
\end{equation}
The equation for the vertex $\Gamma^{[12]}$ is,
in a compact matrix notation, given by \cite{br86}
\begin{equation}
     \Gamma^{[12]}=V^{[12]}+V^{[12]}(gg)\Gamma^{[12]},
     \label{eq:schematic12}
\end{equation}
where $g$ is the one-body Green's function representing 
the intermediate states. 
The diagrammatic expression for this equation is
given in Fig.\ \ref{fig:selfcons12}. 
The expression of Eq.\ (\ref{eq:schematic12}) is known as the Feynman-Galitskii
equation. This equation is normally solved iteratively. 
In the first iteration 
the irreducible
vertex $V^{[12]}$ is then often chosen as the bare $NN$ 
interaction. This interaction is then typically assumed to be energy
independent and we can drop the $s$ dependence of $V^{[12]}$. Moreover,
the single-particle propagators are chosen as the
unperturbed ones. The first iteration of 
Eq.\ (\ref{eq:schematic12}) can then be rewritten as
\begin{equation}
      \Gamma^{[12]}_{1234J}(s) = 
      V^{[12]}_{1234J}+\frac{1}{2}
      \sum_{56}
      V^{[12]}_{1256J}\hat{{\cal G}}^{[12]}
      \Gamma^{[12]}_{5634J}(s),
      \label{eq:first12}
\end{equation}
with the unperturbed particle-particle and hole-hole propagators 
\begin{equation}
    \hat{{\cal G}}^{[12]}=
    \frac{Q^{[12]}_{\mathrm{pp}}}{s-\varepsilon_5-\varepsilon_6+\imath \eta}-
    \frac{Q^{[12]}_{\mathrm{hh}}}{s-\varepsilon_5-\varepsilon_6-\imath \eta},
    \label{eq:paulioperator12}
\end{equation}
which results from the integration over the energy variable 
in the product of the two single-particle
propagators in Eq.\ (\ref{eq:schematic12}). 
The factor $1/2$ follows from 
one of the standard Goldstone-Feynman diagram 
rules \cite{kstop81}, which state
that a factor $1/2$ should be associated with each pair of lines 
which starts at the same interaction vertex and ends at the same
interaction vertex. 
The reader should note that the intermediate states $56$
can represent a two-particle state or a two-hole state.
In Eq.\ (\ref{eq:paulioperator12}) we have assumed unperturbed single-particle
energies. 
In our iterations we will approximate the single-particle energies
with their real part only. Thus, 
the two-particle propagator 
$\hat{{\cal G}}^{[12]}$
with renormalized single-particle energies has the same
form as the unperturbed one.
The operators $Q^{[12]}_{\mathrm{pp}}$ and $Q^{[12]}_{\mathrm{hh}}$ 
ensure that the intermediate states are of two-particle
or two-hole character. 
In order to obtain a self-consistent scheme, Eq.\ (\ref{eq:first12}) 
has also to be accompanied with the 
equation for the single-particle propagators $g$ 
given by Dyson's equation
\begin{equation}
    g=g_0-g_0\Sigma g,
    \label{eq:dyson12}
\end{equation}
with $g_0$ being the unperturbed single-particle 
propagator and $\Sigma$ the self-energy. We will however defer a discussion
of these quantities to section \ref{sec:sec5}. Here it will suffice to state
that  the self-energy is related to the vertex 
$\Gamma^{[12]}$ as 
\begin{equation}
      \Sigma \sim g\Gamma.
      \label{eq:sigma12}
\end{equation}
The similarity sign is meant to indicate that, although being formally
correct, great care has to be exercised in order not to double-count
contributions to the self-energy \cite{jls82}. 
The set of equations for the vertex function and the self-energy 
is shown pictorially in Fig.\ \ref{fig:selfcons12}.
Assume now that we have performed the first iteration. The question which now
arises  is whether the obtained vertex $\Gamma^{[12]}$ from the solution of
Eq.\ (\ref{eq:first12}) should replace the bare vertex $V^{[12]}$
in the next iteration. Before answering this question, let us give some examples
of diagrams which can be generated from the first iteration. 
\begin{figure}[hbtp]
      \setlength{\unitlength}{1mm}
      \begin{picture}(100,70)
      \put(35,0){\epsfxsize=8cm \epsfbox{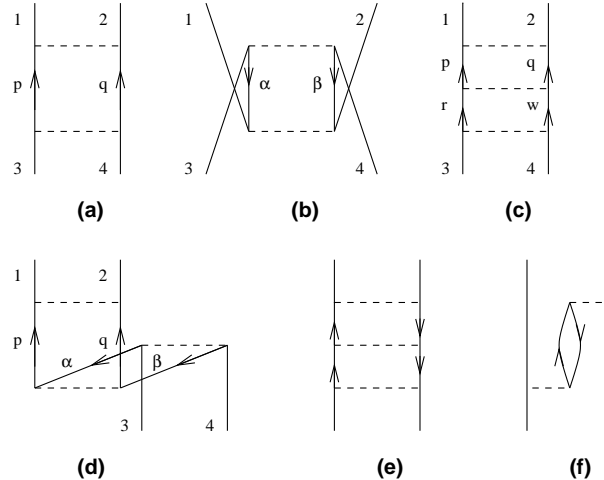}}
      \end{picture}
      \caption{Diagrams (a)-(d) give examples of 
               diagrams which are summed up by  
               the use of Eq.\ (\protect{\ref{eq:schematic12}}).
               Diagrams (e) and (f) are examples of core-polarization
               terms which are not generated by the $[12]$ channel.}
      \label{fig:gamma12}
\end{figure}
These contributions
are given by e.g., diagrams (a)-(d) in Fig.\ \ref{fig:gamma12}. 
Diagrams (a) and (b) are examples of contributions to second order 
in perturbation theory, while diagrams (c) and (d) are higher
order terms. Diagrams (e) and (f) are higher-order core-polarization terms,
which can e.g., be generated through the solution of the equations for
the $[13]$ and $[14]$ channels discussed in the next section.
If we were to include diagrams (a)-(d) in the definition of the bare
vertex in our next iteration,
we are prone to double-count, since such contributions are generated
once again. Diagrams which contain particle-hole intermediate state are however
not generated by the solution of Eq.\ (\ref{eq:first12}).
We need therefore to define the vertex $V^{[12]}$ used in every iteration
to be the sum of diagrams which are irreducible in the $[12]$ channel.
With irreducible we will mean all diagrams which cannot be reduced to
a piece containing the particle states $12$  entering or leaving 
the same interaction vertex 
and another   part containing the states $34$ at the same interaction
vertex by cutting two internal lines.
Clearly, if we cut diagrams (a) and (b) we are just left with two bare 
interaction vertices. Similarly, cutting two lines of an intermediate 
state in diagrams (c) and (d) leaves us with two second-order terms
of the type (a) and (b) and one bare interaction. 
Diagrams (e) and (f) are however examples of diagrams which are
irreducible in the $[12]$ channel. Diagram (e) is 
irreducible in the $[13]$ channel, but not in the $[14]$ channel.
Similarly, diagram (g) is reducible in the $[13]$ channel and
irreducible in the $[14]$ channel. 
This means that, unless we solve equations similar to Eq.\ (\ref{eq:first12})
in the $[13]$ channel and $[14]$ channels as well, changes 
from further iterations
of Eq.\ (\ref{eq:first12}) will only come from 
the single-particle terms defined
by Dyson's equation in Eq.\ (\ref{eq:dyson12}).

In the remaining part of this section, we will try to delineate ways 
of solving the above equations, and discuss possible approximations,
their merits and faults. First of all, we will reduce
the propagator of  Eq.\ (\ref{eq:paulioperator12}) to only include
particle-particle intermediate states. This will lead us to the familiar
$G$-matrix in nuclear many-body theory. 
Based on the $G$-matrix, we will construct effective interactions
through perturbative summations. Applications of such 
effective interacions to selected nuclei will then be discussed.
Thereafter, we will try to
account for hole-hole contributions and end this section with a 
discussion on self-consistent determinations of the single-particle 
energies through the solution of Dyson's equation.
 
\subsection{The $G$-matrix}

In nuclear structure and nuclear matter calculations one has to
face the problem that any realistic $NN$ interaction $V$
exhibits a strong short-range repulsion, which in turn makes
a perturbative treatment of the nuclear many-body problem
prohibitive. If the interaction has a so-called hard core,
the matrix elements of such an interaction $\bra{\psi}V\ket{\psi}$
evaluated
for an uncorrelated two-body wave function $\psi (r)$ diverge,
since the uncorrelated wave function is different from zero also for
relative distances $r$ smaller than the hard-core radius. Similarly,
even if one uses interactions with softer cores, the matrix elements of the
interaction become very large at short distances.
The above problem was however overcome by
introducing the reaction matrix $G^{[12]}$ (displayed by  the summation
of ladder type of diagrams in Fig.\ \ref{fig:gamma12}
with particle-particle intermediate states only),
accounting thereby for short-range two-nucleon correlations.
The $G^{[12]}$-matrix represents just a subset to the solution
of the equations for the interaction $\Gamma^{[12]}$ in the $[12]$ channel,
we have clearly neglected the possibility of having intermediate states
which are of the hole-hole type.  
The matrix elements of the
interaction $V^{[12]}$ then become
\begin{equation}
\bra{\psi}G^{[12]}\ket{\psi} =\bra{\psi}V^{[12]}\ket{\Psi}
\end{equation}
where $\Psi$ is now the correlated wave function
containing selected correlation from the excluded space.
By accounting for these
correlations in the two-body wave functon $\Psi$, the matrix elements of
the interaction become finite, even for a hard-core interaction $V$. Moreover,
as will be discussed below, compared with the uncorrelated
wave function, the correlated wave function enhances the
matrix elements of $V$ at distances for which the interaction is
attractive.
The type of correlations which typically are included in the evaluation
of the $G^{[12]}$-matrix are those of the two-particle type.
If we label the operator $Q$ in this case by $Q^{[12]}_{\mathrm{pp}}$, we can write
the integral equation for the $G$-matrix as 
\begin{equation}
   G^{[12]}(s)=V^{[12]}+V^{[12]}\frac{Q^{[12]}_{\mathrm{pp}}}
           {s -H_0+\imath \eta}G^{[12]}(s),
   \label{eq:g1}
\end{equation}
implicitely assuming that $\lim \eta \rightarrow \infty$. 
The variable $s$ represents normally the unperturbed 
energy of the incoming two-particle state. We will suppress $\imath \eta$ in the 
following equations. Moreover, since one is often interested only in the 
$G^{[12]}$-matrix
for negative starting energies, the $G^{[12]}$-matrix commonly used in studies
of effective interactions has no divergencies.

We can also write  
\begin{equation}
      G^{[12]}(s )=V^{[12]}+V^{[12]}Q^{[12]}_{\mathrm{pp}}
      \frac{1}{s -Q^{[12]}_{\mathrm{pp}}H_0Q^{[12]}_{\mathrm{pp}}}
      Q^{[12]}_{\mathrm{pp}}
      G^{[12]}(s ).
      \label{eq:g2}
\end{equation}
The former equation applies if the Pauli operator $Q^{[12]}_{\mathrm{pp}}$  commutes
with the unperturbed hamiltonian $H_0$, whereas the latter is
needed if $[H_0,Q^{[12]}_{\mathrm{pp}}]\neq 0$.
Similarly, the correlated wave function $\Psi$
is given as
\begin{equation}
    \ket{\Psi}=\ket{\psi}+
    \frac{Q^{[12]}_{\mathrm{pp}}}{s - H_0}G^{[12]}\ket{\psi},
    \label{eq:wave}
\end{equation}
or
\begin{equation}
   \ket{\Psi}=\ket{\psi}+Q^{[12]}_{\mathrm{pp}}\frac{1}
    {s - Q^{[12]}_{\mathrm{pp}}H_0Q^{[12]}_{\mathrm{pp}}}
    Q^{[12]}_{\mathrm{pp}}G^{[12]}\ket{\psi}.
\end{equation}

In order to evaluate the $G^{[12]}$-matrix for finite nuclei,
we define first a useful identity following Bethe, Brandow and
Petschek \cite{bbp63}. Suppose we have two
different $G$-matrices\footnote{For notational economy, 
we drop the superscript $^{[12]}$. Furthermore, 
in the subsequent discussion in 
this subsection it is understood that all operators $Q$ 
refer to particle-particle intermediate states only. The subscript
$\mathrm{pp}$ is also dropped.}, defined by
\begin{equation}
    G_1=V_1+V_1\frac{Q_1}{e_1}G_1,
\end{equation}
and
\begin{equation}
    G_2=V_2+V_2\frac{Q_2}{e_2}G_2,
\end{equation}
where $Q_1/e_1$ and $Q_2/e_2$ are the propagators of
either Eq.\ (\ref{eq:g1}) or Eq.\ (\ref{eq:g2}). $G_1$ and $G_2$
are two different $G$-matrices having two different interactions
and/or different propagators. We aim at an identity
which will enable us to calculate $G_1$ in terms of $G_2$,
or vice versa.
Defining the wave operators
\begin{equation}
    \Omega_1=1+\frac{Q_1}{e_1}G_1,
\end{equation}
and
\begin{equation}
    \Omega_2=1+\frac{Q_2}{e_2}G_2,
\end{equation}
we can rewrite the above $G$-matrices as
\begin{equation}
    G_1=V_1\Omega_1,
    \label{eq:omega1}
\end{equation}
and
\begin{equation}
    G_2=V_2\Omega_2.
    \label{eq:omega2}
\end{equation}
Using these relations, we rewrite $G_1$ as
\begin{eqnarray}
   G_1=&G_1 -{\displaystyle 
         G_2^{\dagger}\left(\Omega_1-1-\frac{Q_1}{e_1}G_1\right)
        +\left(\Omega_2^{\dagger}-1-G_2^{\dagger}\frac{Q_2}{e_2}\right)G_1} 
         \nonumber \\
       =&{\displaystyle G_2^{\dagger} +G_2^{\dagger}\left(\frac{Q_1}{e_1}-
        \frac{Q_2}{e_2}\right)G_1
        +\Omega_2^{\dagger}G_1 -G_2^{\dagger}\Omega_1},
\end{eqnarray}
and using Eqs.\ (\ref{eq:omega1}) and (\ref{eq:omega2}) we obtain
the identity
\begin{equation}
        G_1=G_2^{\dagger} +G_2^{\dagger}
        \left(\frac{Q_1}{e_1}-\frac{Q_2}{e_2}\right)G_1
        +\Omega_2^{\dagger}(V_1-V_2)\Omega_1.
        \label{eq:gidentity}
\end{equation}
The second term on the rhs.\ is called the propagator-correction term;
it vanishes if $G_1$ and $G_2$ have the same propagators. The third term
is often referred to as the potential-correction term, and it disappears
if $G_1$ and $G_2$  have the same potentials.
\begin{figure}[hbtp]
    \setlength{\unitlength}{1mm}
    \begin{picture}(100,60)
    \put(35,0){\epsfxsize=8cm \epsfbox{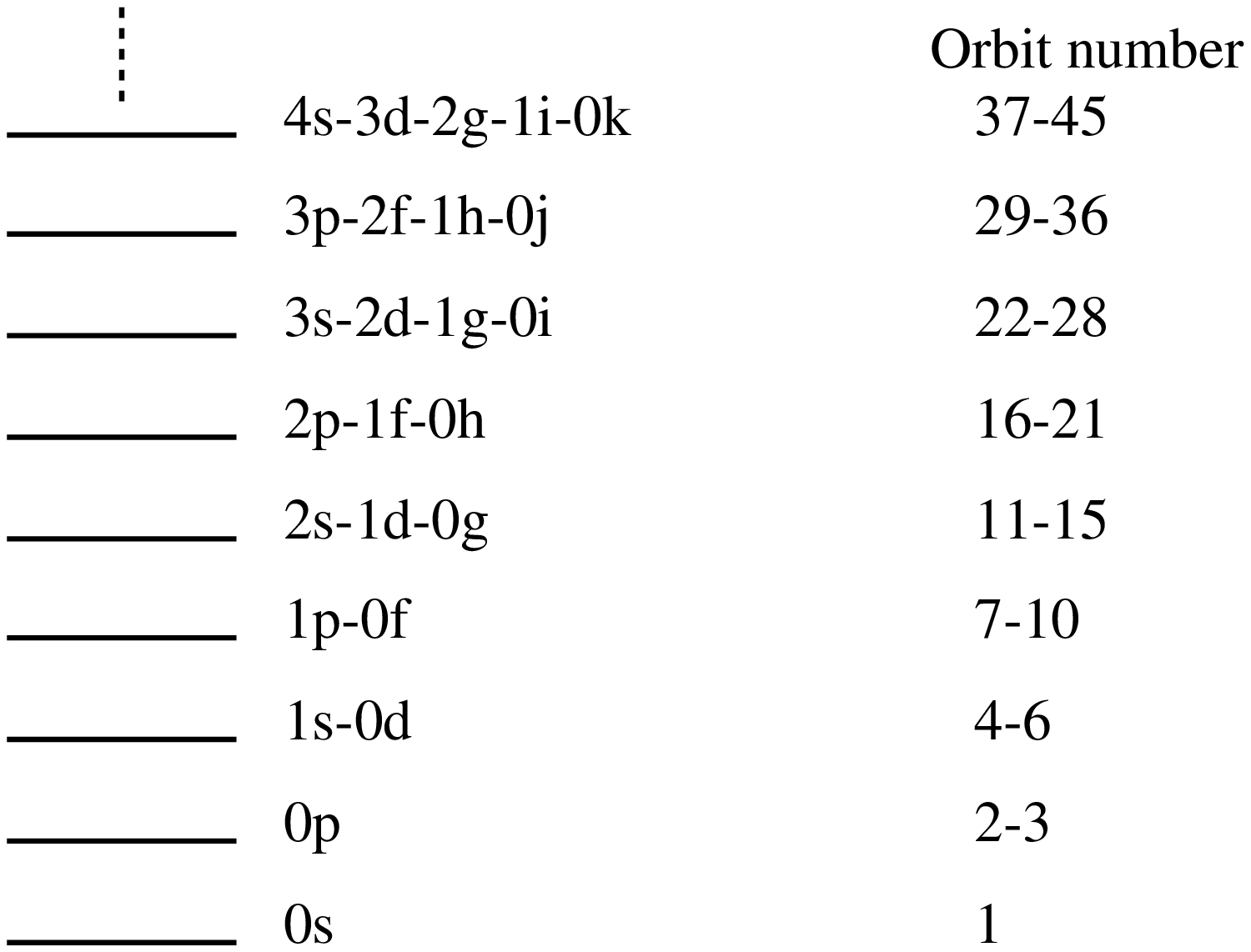}}
    \end{picture}
\caption{Classification of harmonic oscillator single-particle
orbits.}
\label{fig:orbits}
\end{figure}
The reader may now ask what is the advantage of the above identity. If
we assume that by some physical reasoning we are able to calculate
$G_2$ and that the expression for $G_2$ can be calculated
easily, and further that $G_2$ is a good approximation
to the original $G$-matrix, then we can use the above identity to
perform a perturbative calculation of $G_1$ in terms of $G_2$.

Before we proceed in detailing the calculation of the $G$-matrix
appropriate for finite nuclei, certain approximations need be explained.

\begin{figure}[hbtp]
      \setlength{\unitlength}{1mm}
      \begin{picture}(100,60)
      \put(35,0){\epsfxsize=7cm \epsfbox{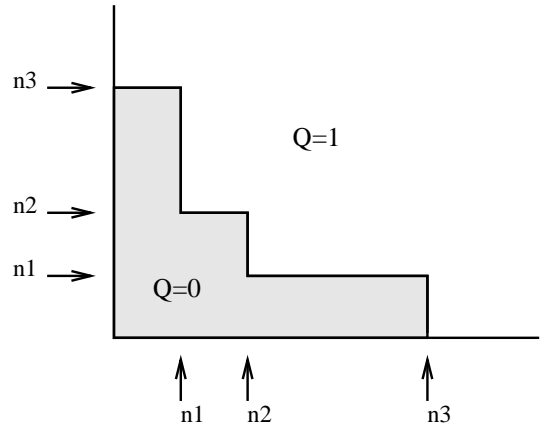}}
      \end{picture}
\caption{Definition of the $P$ (shaded area) and $Q$ operators
appropriate for the definition of the $G$-matrix and the effective
interaction. Outside the shaded area limited by the boundaries $n_1$,
$n_2$ and $n_3$ $P=0$ and $Q=1$.}
\label{fig:qoperat}
\end{figure}

\begin{figure}[hbtp]
      \setlength{\unitlength}{1mm}
      \begin{picture}(100,50)
       \put(35,5){\epsfxsize=8cm \epsfbox{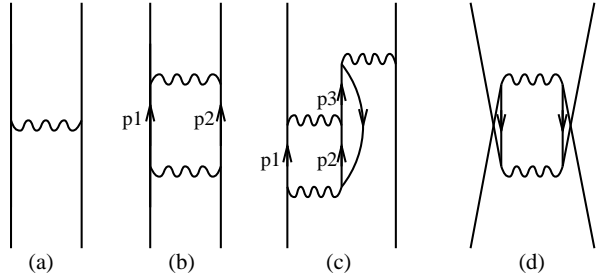}}
      \end{picture}
      \caption{Examples of diagrams which may define the effective valence space
         interaction. The wavy line is the $G$-matrix.}
\label{fig:qboxexam1}
\end{figure}

As discussed above, the philosophy behind perturbation theory is
to reduce the intractable full Hilbert space problem to one which
can be solved within a physically motivated model space, defined by the
operator $P$. The excluded degrees of freedom are represented by the
projection operator $Q$. The definition of these operators is connected
with the nuclear system and the perturbative expansions discussed
in section \ref{sec:sec2}. Consider the evaluation of the effective interaction
needed in calculations of the low-lying states of $^{18}$O. From
experimental data and theoretical calculations the belief is that
several properties of this nucleus can be described by a model
space consisting of a closed $^{16}$O core (consisting of the filled
$0s$- and $0p$-shells) and two valence neutrons
in the $1s0d$-shell. In Fig.\ \ref{fig:orbits} we exhibit this division
in terms of h.o.~sp orbits.
The active sp states in the $1s0d$-shell are then given by the  $0d_{5/2}$, 
$0d_{3/2}$ and $1s_{1/2}$ orbits, labels $4-6$ in Fig.\ \ref{fig:orbits}.
The remaining states enter the definition of
$Q$. Once we have defined $P$ and $Q$ we proceed in constructing the $G$-matrix
and the corresponding perturbative expansion in terms of the $G$-matrix. 
There are however several ways of choosing $Q$. A common procedure is to
specify the boundaries of $Q$ by three numbers, $n_1$, $n_2$ and $n_3$, explained
in Fig.\ \ref{fig:qoperat}.
For $^{18}$O we would choose $(n_1=3,n_2=6,n_3=\infty)$. 
Our choice of 
$P$-space implies that the single-particle states outside the model space
start from the
$1p0f$-shell (numbers 7--10 in Fig.\ \ref{fig:orbits}), and orbits 1, 2
and 3 are hole states. Stated differently, this means that $Q$
is constructed so as to prevent scattering into intermediate 
two-particle states 
with one particle in the $0s$- or $0p$-shells or both particles
in the $1s0d$-shell. This definition of the $Q$-space influences the determination
of the effective shell-model interaction. Consider the diagrams displayed
in Fig.\ \ref{fig:qboxexam1}.
Diagram (a) of this figure is just the $G$-matrix and is allowed in the definition
of the $\hat{Q}$-box. With our choice $(n_1=3,n_2=6,n_3=\infty)$, diagram (b) is not
allowed since the intermediate state consists of passive particle
states  and is already included in the evaluation of the $G$-matrix. Similarly,
diagram (c) is also not allowed whereas diagram (d) is allowed. Now an important
subtlety arises. If we evaluate the $G$-matrix with the boundaries
$(n_1=3,n_2=10,n_3=\infty)$, and define the $P$-space of {\em 
the effective interaction}
by including orbits 4 to 6 only, then diagrams (b) and (c)
are allowed if $7\leq p_1 , p_2 \leq 10$
In this way we allow for 
intermediate two-particle states as well with orbits outside the 
model-space of the effective interaction. The reader should notice the above
differences, i.e.\ that the $Q$-space defining the $G$-matrix and 
$H_{\mathrm{eff}}$
may differ. 
In order to calculate the $G$-matrix we will henceforth employ a 
so-called double-partitioned scheme. 
Let us be more specific and detail this double-partitioned procedure.
We define first a reference $G$-matrix $\tilde{G}$ 
in terms of plane wave intermediate states only, meaning that $H_0$ is
replaced by the kinetic energy operator $T$ only
while $G$ has harmonic oscillator intermediate states (this is one
possible choice for $U$). We divide the exclusion operator
into two parts, one which represents the low-lying states $Q_l$ and
one which accounts for high-lying states $Q_h$, viz.\
\[
    Q=Q_l+Q_h=Q_l+\tilde{Q}.
\]
If we consider $^{18}$O as our pilot nucleus, we may define $Q_l$ to consist
of the sp orbits of the $1p0f$-shell, orbits $7-10$ in Fig.\ \ref{fig:orbits},
described by h.o.\ states. $Q_h$ represents then the remaining orthogonalized
intermediate states.
Using the identity of Bethe, Brandow and Petschek \cite{bbp63} of
Eq.\ (\ref{eq:gidentity}) we 
can first set up $\tilde{G}$ as 
\begin{equation}
     \tilde{G}=V+V\frac{\tilde{Q}}{s -T}\tilde{G},
\label{eq:gfinite}
\end{equation}
and  express $G$ in terms of $\tilde{G}$ as
\begin{equation}
        G=\tilde{G} +\tilde{G}
        \left(\frac{Q_l}{s -H_0}\right)G,
        \label{eq:gidfinite}
\end{equation}
and we have assumed that $\tilde{G}$ is hermitian and that $[Q_l,H_0]=0$.
Thus, we first calculate
a ``reference'' $G$-matrix ($\tilde{G}$ in our case), and then insert this
in the expression for the full $G$-matrix. The novelty here is that
we are able to calculate $\tilde{G}$ exactly through operator relations
to be discussed below. In passing we note that $G$ depends significantly 
on the choice of $H_0$, though the low-lying intermediate states
are believed to be fairly well represented by h.o.\ states. 
Also, the authors of ref.\ \cite{kkko76} demonstrate that low-lying
intermediate states are not so important in $G$-matrix calculations,
being consistent with the short-range nature of the $NN$ interaction.
Since we let $Q_l$ to be defined by the orbits of the $1p0f$-shell,
and the energy difference between two particles in the 
$sd$-shell and $pf$ shell is of the order $-14$ MeV, we can treat 
$G$ as a perturbation expansion in $\tilde{G}$.
Eq.\ (\ref{eq:gidfinite}) can then be written as 
\begin{equation}
        G=\tilde{G} +\tilde{G}
        \left(\frac{Q_l}{s -H_0}\right)\tilde{G}
        +\tilde{G}
        \left(\frac{Q_l}{s -H_0}\right)\tilde{G}
        \left(\frac{Q_l}{s -H_0}\right)\tilde{G} +\dots
\end{equation}
The only intermediate states are those defined by the $1p0f$-shell.
The second term on the rhs.\ is nothing but the second-order 
particle-particle ladder. The third term is then the third-order ladder
diagram in terms of
$\tilde{G}$. As shown by the authors of ref.\ \cite{kkko76}, the inclusion
of the second-order particle-particle diagram in the evaluation
of the $\hat{Q}$-box, represents a good approximation. 
The unsettled problem is however how to define
the boundary between 
$Q_l$ and $Q_h$. 

Now we will discuss how to compute $\tilde{G}$. 
One can solve the equation for the $G$-matrix
for finite nuclei by employing
a formally
exact technique for handling $\tilde{Q}$
discussed in e.g., Ref.\ \cite{kkko76}.
Using the matrix identity
\begin{equation}
  \tilde{Q}\frac{1}{\tilde{Q}A\tilde{Q}}
  \tilde{Q}=\frac{1}{A}-
   \frac{1}{A}\tilde{P}\frac{1}{\tilde{P}A^{-1}\tilde{P}}\tilde{P}\frac{1}{A},
   \label{eq:matrix_relation_q}
\end{equation}
with $A=s -T$, to rewrite Eq.\ (\ref{eq:gfinite}) as\footnote{We will omit the
label $\tilde{G}$ for the $G$-matrix for finite nuclei, however it is
understood that the $G$-matrix for finite nuclei is calculated according
to Eq.\ (\ref{eq:gfinite}) This means that we have to 
include the particle-particle ladder diagrams in the 
$\hat{Q}$-box. }
\begin{equation}
   G = G_{F} +\Delta G,\label{eq:gmod}
\end{equation}
where $G_{F}$ is the free $G$-matrix defined as
\begin{equation}
   G_{F}=V+V\frac{1}{s - T}G_{F}. \label{eq:freeg}
\end{equation}
The term $\Delta G$ is a correction term defined entirely within the
model space $\tilde{P}$ and given by
\begin{equation}
   \Delta G =-V\frac{1}{A}\tilde{P}
   \frac{1}{\tilde{P}A^{-1}\tilde{P}}\tilde{P}\frac{1}{A}V.
\end{equation}
Employing the definition for the free $G$-matrix of Eq.\ (\ref{eq:freeg}),
one can rewrite the latter equation as
\begin{equation}
  \Delta G =-G_{F}\frac{1}{e}\tilde{P}
  \frac{1}{\tilde{P}(e^{-1}+e^{-1}G_{F}e^{-1})
  \tilde{P}}\tilde{P}\frac{1}{e}G_F,
\end{equation}
with $e=s -T$.
We see then that the $G$-matrix for finite nuclei
is expressed as the sum of two
terms; the first term is the free $G$-matrix with no Pauli corrections
included, while the second term accounts for medium modifications
due to the Pauli principle. The second term can easily
be obtained by some simple matrix operations involving
the model-space matrix $\tilde{P}$ only.
However, the second term is a function of the variable 
$n_3$. The convergence in terms of $n_3$ was discussed ad extenso
in Ref.\ \cite{hko95} and we refer the reader to that work.
The equation for the free matrix $G_F$ is solved in momentum space in the 
relative and centre of mass system and thereafter transformed to the 
relevant expression in terms of harmonic ocillator single-particle
wavefunctions in the laboratory system. This yields final
matrix elements of the type
\begin{equation}
  \bra{(ab)J}G\ket{(cd)J}
\end{equation}
where $G$ is the given by the sum $G = G_{F} +\Delta G$.
The label $a$ represents here all the single particle quantum numbers
$n_{a}l_{a}j_{a}$.

\subsection{Folded diagrams and the effective valence space interaction}

Here we discuss further classes of diagrams
which can be included in the evaluation of effective interactions
for the shell model.
Especially, we will focus on the summations of so-called folded
diagrams.

One way of obtaining the wave operator 
$\Omega$ is through the generalized Bloch
equation given by Lindgren and Morrison \cite{lm85}
\begin{equation}
[\Omega, H_0]P=QH_1\Omega P-\chi PH_1\Omega P,
\label{eq:lind}
\end{equation}
which offers a suitable way of generating the RS perturbation expansion.
Writing Eq.\ (\ref{eq:lind}) in terms of $\Omega^{(n)}$ we have
\begin{equation}
[\Omega^{(1)}, H_0]P=QH_1P,
\end{equation}
\begin{equation}
[\Omega^{(2)}, H_0]P=QH_1\Omega^{(1)} P- \Omega^{(1)} PH_1P,
\end{equation}
and so forth,  which can be generalized to
\begin{equation}
[\Omega^{(n)}, H_0]P=QH_1\Omega^{(n-1)} P- \sum_{m=1}^{n-1}
\Omega^{(n-m)} PH_1\Omega^{(m-1)}P.
\end{equation}
The effective interaction to a given order can then be obtained from
$\Omega^{(n)}$, see \cite{lm85}.
Another possibility is obvioulsy the coupled-cluster method discussed
elewhere in this volume.

Here we will assume that we can start with a given
approximation to $\Omega$, and through an iterative scheme generate
higher order terms. Such schemes will in general differ from the
order-by-order scheme of Eq.\ (\ref{eq:lind}).  Two such iterative
schemes were derived  by Lee and Suzuki \cite{ls80}. We will focus
on the folded diagram method of Kuo and co-workers \cite{ko90}.

\begin{figure}[hbtp]
      \setlength{\unitlength}{1mm}
      \begin{picture}(100,40)
      \put(35,0){\epsfxsize=8cm \epsfbox{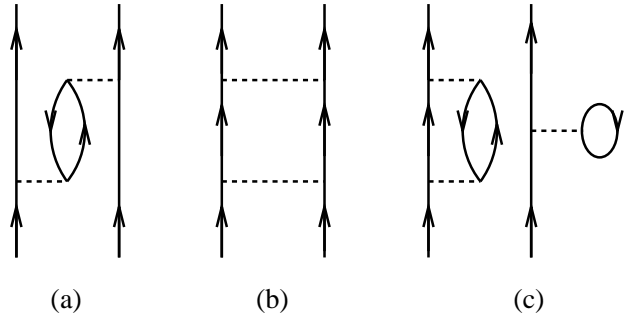}}
      \end{picture}
\caption{Different types of valence-linked diagrams. Diagram (a)
is irreducible and connected, (b) is reducible, while (c) is irreducible
and disconnected.}
\label{fig:diagsexam}
\end{figure}

Having defined the wave operator $\Omega = 1 +\chi$ (note that $
\Omega^{-1}=1-\chi$) with
$\chi$ given by Eq.\ (\ref{eq:chi2}) we can obtain
\begin{equation}
     QHP-\chi HP +QH\chi - \chi H\chi = 0. \label{eq:basic}
\end{equation}
This is the basic equation to which a solution to $\chi$ is
to be sought.
If we choose to  work with a degenerate model space we define
\[
   PH_0 P = s P,
\]
where $s$ is the unperturbed model space eigenvalue
(or starting energy) in the degenerate case,
such that Eq.\ (\ref{eq:basic}) reads in a slightly modified form
($H=H_0 + H_1$)
\[
    (s -QH_0 Q -QH_1 Q)\chi = QH_1 P -\chi PH_1 P -\chi PH_1 Q\chi,
\]
which yields the following equation for $\chi$
\begin{equation}
    \chi = \frac{1}{s - QHQ}QH_1 P -\frac{1}{s -QHQ}\chi\left(PH_1 P +
    PH_1 Q\chi P\right).\label{eq:chi3}
\end{equation}
Observing that  the $P$-space effective hamiltonian is given as
\[
     H_{\mathrm{eff}}= PHP+PH\chi=PH_0 P + V_{\mathrm{eff}}(\chi),
\]
with $V_{\mathrm{eff}}(\chi)= PH_1 P + PH_1Q\chi P$, Eq. (\ref{eq:chi3}) becomes
\begin{equation}
     \chi = \frac{1}{s - QHQ}QH_1 P -\frac{1}{s -QHQ}
     \chi V_{\mathrm{eff}}(\chi ).
     \label{eq:chi4}
\end{equation}
Now we find it convenient to introduce the so-called $\hat{Q}$-box,
defined as
\begin{equation}
     \hat{Q}(s)=PH_1 P + PH_1 Q\frac{1}{s - QHQ}
      QH_1 P.\label{eq:qbox}
\end{equation}
The $\hat{Q}$-box is made up of non-folded diagrams which are irreducible
and valence linked. A diagram is said to be irreducible if between each pair
of vertices there is at least one hole state or a particle state outside
the model space. In a valence-linked diagram the interactions are linked
(via fermion lines) to at least one valence line. Note that a valence-linked
diagram can be either connected (consisting of a single piece) or
disconnected. In the final expansion including folded diagrams as well, the
disconnected diagrams are found to cancel out \cite{ko90}. 
This corresponds to the cancellation of unlinked diagrams
of the Goldstone expansion.
We illustrate
these definitions by the diagrams shown in Fig.\ 
\ref{fig:diagsexam}.
Diagram (a) is irreducible, valence linked and connected, 
while (b) is reducible since 
the intermediate particle states belong to the model space. 
Diagram (c) is irreducible, valence linked and disconnected. 
It is worth noting that general form of the $\hat{Q}$-box
is the same as that of the $G$-matrix, or the equations
of the $[12]$ channel or those of the $[13]$ and $[14]$
channels to be discussed in section \ref{sec:sec4}.
In Ref.\ \cite{hko95}, the $\hat{Q}$-box was defined
to be the sum all diagrams to third order in the $G$-matrix.

Multiplying both sides of Eq.\ (\ref{eq:chi4}) with $PH_1$ and
adding $PH_1 P$ to both sides we get
\[
    PH_1 P + PH_1 \chi =
    PH_1 P + PH_1 Q\frac{1}{s - QHQ}QH_1 P -
    PH_1 \frac{1}{s -QHQ}\chi V_{\mathrm{eff}}(\chi ),
\]
which gives
\begin{equation}
     V_{\mathrm{eff}}(\chi )=\hat{Q}(s)-
     PH_1 \frac{1}{s -QHQ}\chi V_{\mathrm{eff}}(\chi ).
     \label{eq:veff}
\end{equation}

There are several ways to solve Eq.\ (\ref{eq:veff}). The idea is
to set up an iteration scheme where we determine $\chi_n$ and
thus $V_{\mathrm{eff}}(\chi_n )$ from 
$\chi_{n-1}$ and $V_{\mathrm{eff}}(\chi_{n-1})$.
For the mere sake of simplicity we write 
$V_{\mathrm{eff}}^{(n)}=V_{\mathrm{eff}}(\chi_{n})$.

Let us write Eq.\ (\ref{eq:veff}) as
\[
   V_{\mathrm{eff}}^{(n)}=\hat{Q}(s)-
   PH_1 \frac{1}{s -QHQ}\chi_n V_{\mathrm{eff}}^{(n-1)}.
\]
The solution to this equation can be shown to be \cite{ls80}
\begin{equation}
    V_{\mathrm{eff}}^{(n)}=\hat{Q}+{\displaystyle\sum_{m=1}^{\infty}}
    \frac{1}{m!}\frac{d^m\hat{Q}}{ds^m}\left\{
    V_{\mathrm{eff}}^{(n-1)}\right\}^m . 
    \label{eq:fd}
\end{equation}
Observe also that the
effective interaction is $V_{\mathrm{eff}}^{(n)}$ 
is evaluated at a given model space energy
$s$. If
$V_{\mathrm{eff}}^{(n)}=V_{\mathrm{eff}}^{(n-1)}$, the iteration is said to
converge. In the limiting case $n\rightarrow \infty$, the
solution $V_{\mathrm{eff}}^{(\infty)}$ agrees with the formal solution of
Brandow
\cite{brandow67} and Des Cloizeaux \cite{des}
\begin{equation}
    V_{\mathrm{eff}}^{(\infty)}=\sum_{m=0}^{\infty}\frac{1}{m!}
    \frac{d^{m}\hat{Q}}{ds^{m}}\left\{
    V_{\mathrm{eff}}^{(\infty)}\right\}^{m}.\label{eq:pert}
\end{equation}

Alternatively, we can generate the contribution
from $n$ folds the following way. In
an $n$-folded $\hat{Q}$-box there are of course $n+1$ $\hat{Q}$-boxes. The 
general expression for an $n$-folded $\hat{Q}$-box is then
\begin{equation}
        \hat{Q}   -\hat{Q}\int\hat{Q} 
    +\hat{Q}\int\hat{Q}\int\hat{Q} -\dots=
    {\displaystyle\sum_{m_1m_2\dots m_n}}
    \frac{1}{m_1!}\frac{d^{m_1}\hat{Q}}{ds^{m_1}}P
    \frac{1}{m_2!}\frac{d^{m_2}\hat{Q}}{ds^{m_2}}P
    \dots
    \frac{1}{m_n!}\frac{d^{m_n}\hat{Q}}{ds^{m_n}}P\hat{Q},
\label{eq:fdfinal}
\end{equation}
where we have the constraints
\[
  m_1+m_2+\dots m_n=n,
\]
\[
m_1\geq 1,
\]
\[
m_2, m_3, \dots m_n \geq 0,
\]
and
\[
m_k \leq n-k+1.
\]
The last restriction follows from the fact that there are only
$n-k+1$ $\hat{Q}$-boxes to the right of $k^{\mathrm{th}}$ $\hat{Q}$-box.
Thus, it can at most be differentiated $n-k+1$ times. 
We have inserted the model-space projection operator in the above
expression, in order to emphasize that folded diagrams have as intermediate
states between successive $\hat{Q}$-boxes 
only model-space states. Therefore, the sum in Eq.\ (\ref{eq:fdfinal})
includes a sum over all model-space states with the same quantum 
numbers such as isospin and total angular momentum. It is understood
that the $\hat{Q}$-box  and its derivatives are evaluated at the
same starting energy, which should correspond to the unperturbed 
energy of the model-space state.     
It is then straightforward to recast Eq.\ (\ref{eq:fdfinal}) 
into the form of Eq.\ (\ref{eq:fd}).

Note that although $\hat{Q}$ and its derivatives contain disconnected
diagrams, such diagrams cancel exactly in each order \cite{ko90}, thus
yielding a fully connected expansion in Eq.\ (\ref{eq:fd}).
However, in order to achieve this, disconnected diagrams have 
to be included in the definition of the $\hat{Q}$-box. An example
is given by diagram (c) Fig.\ \ref{fig:diagsexam}. 
Such a diagram will generate a contribution to 
the first fold $\frac{d\hat{Q}}{ds}\hat{Q}=-\hat{Q}\int\hat{Q} $
which cancels exactly diagram (c) when all time-ordered
contributions to this diagram are accounted for, see Ref.\ \cite{ko90}
for more details. 
It is moreover important to note in connection with the above 
expansion, that a term like
$F_1= \hat{Q}_1 \hat{Q}$ actually means $P\hat{Q}_1 P\hat{Q}P$ since
the $\hat{Q}$-box is defined in the model space only. Here we have defined
$\hat{Q}_{m}=\frac{1}{m!}\frac{d^{m}\hat{Q}}
{ds^{m}}$.
Due to this structure, only so-called folded diagrams
contain $P$-space intermediate states.

The folded diagram expansion discussed above yields however
a non-hermitian effective interaction. This may happen even at the 
level of the $G$-matrix, or any of the effective interactions
we will derive in this work, from Parquet theory to
the order-by-order perturbation expansion.

We will therefore end this section with a way to cure
this non-hermiticity. 
A hermitian effective interaction has recently been derived by 
Ellis, Kuo, Suzuki and co-workers \cite{so95,so84,kehlsok93} through the
following steps\footnote{The reader who wishes more details can consult
Refs.\ \cite{so95,kehlsok93}.}.  
To obtain a hermitian effective interaction, 
let us define 
a model-space eigenstate
$\left | b_{\lambda}\right\rangle$ with eigenvalue $\lambda$ as
\begin{equation}
     \left | b_{\lambda}\right\rangle=\sum_{\alpha =1}^{D}
     b_{\alpha}^{(\lambda )}\left | \psi_{\alpha}\right\rangle
\end{equation}
and the biorthogonal wave function
\begin{equation}
     \left | \overline{b}_{\lambda}\right\rangle=\sum_{\alpha=1}^{D}
      \overline{b}_{\alpha}^{(\lambda )}
     \left | \overline{\psi}_{\alpha}\right\rangle,
\end{equation}
such that
\begin{equation}
     {\left\langle \overline{b}_{\lambda} | b_{\mu} \right\rangle}=
     \delta_{\lambda\mu}.
\end{equation}
The model-space eigenvalue problem can be written 
in terms of the above non-hermitian effective interaction unperturbed
wave functions
\begin{equation}
     {\displaystyle
     \sum_{\gamma =1}^{D}b_l^{(\lambda )}\left\langle \psi_{\sigma}\right |
     H_0+V+VQ\chi\left | \psi_{\gamma}\right\rangle}  =
     E_{\lambda}b_{\sigma}^{(\lambda )}.
\end{equation}
The exact wave function expressed in terms of the correlation 
operator is
\begin{equation}
      \left | \Psi_{\lambda}\right\rangle=
      ({\bf 1}+\chi)\left | \psi_{\lambda}\right\rangle.
\end{equation}
The part $\chi\left | \psi_{\lambda}\right\rangle$ can be expressed in terms
of the time-development operator or 
using the time-independent formalism as
\begin{equation}
  \chi\left | \psi_{\lambda}\right\rangle=
  \frac{Q}{E_{\lambda}-QHQ}QVP\left | \psi_{\lambda}\right\rangle,
  \label{eq:newchi}
\end{equation}
where $Q$ is the exclusion operator. Note that this equation is given
in terms of the Brillouin-Wigner perturbation expansion, since
we have the exact energy $E_{\lambda}$ in the denominator.

Using the normalization condition for the true wave function
we obtain
\begin{equation}
  \left\langle \Psi_{\gamma} |\Psi_{\lambda}\right\rangle
  N_{\lambda}\delta_{\lambda\gamma}=
  \left\langle \psi_{\gamma}\right |({\bf 1} +
   \chi^{\dagger}\chi\left | \psi_{\lambda}\right\rangle,
  \label{eq:newnorm}
\end{equation}
where we have used the fact that
$\left\langle \psi_{\gamma}\right | 
\chi\left | \psi_{\lambda}\right\rangle=0$. Recalling that the
time-development operator is hermitian we have that $\chi^{\dagger}\chi$ is
also hermitian. We can then define an orthogonal basis $d$ whose eigenvalue
relation is
\begin{equation}
   \sum_{\alpha}\left\langle \psi_{\beta}\right | \chi^{\dagger}
   \chi\left | \psi_{\alpha}
   \right\rangle d_{\alpha}^{\lambda}=\mu^{2}_{\lambda}
   d_{\beta}^{\lambda},
   \label{eq:newbasis1}
\end{equation}
with eigenvalues greater than $0$.
Using the definition in Eq.\ (\ref{eq:newchi}), we note
that the diagonal element of
\begin{equation}
   \left\langle \psi_{\lambda}\right | \chi^{\dagger}
   \chi\left | \psi_{\lambda}\right\rangle=
   \left\langle \psi_{\lambda}\right | PVQ\frac{1}{(E_{\lambda}-QHQ)^2}QVP
   \left | \psi_{\lambda}\right\rangle,
   \label{eq:chichi}
\end{equation}
which is nothing but the derivative of the $\hat{Q}$-box, with an additional
minus sign. Thus, noting that if $\gamma\neq\lambda$
\begin{equation}
  \left\langle \Psi_{\gamma} |\Psi_{\lambda}\right\rangle=0=
    \left\langle \psi_{\gamma} |\psi_{\lambda}\right\rangle
   +\left\langle \psi_{\gamma}\right |
    \chi^{\dagger}\chi\left | \psi_{\lambda}\right\rangle,
\end{equation}
we can write $\chi^{\dagger}\chi$
in operator form as
\begin{equation}
   \chi^{\dagger}\chi =-\sum_{\alpha}\left | 
    \overline{\psi}_{\alpha}\right\rangle
   \left\langle \psi_{\alpha}\right | 
    \hat{Q}_1(E_{\alpha})\left | \psi_{\alpha}\right\rangle
   \left\langle \overline{\psi}_{\alpha} \right |
   -\sum_{\alpha\neq\beta}\left | \overline{\psi}_{\alpha}\right\rangle
    \left\langle \psi_{\alpha} |\psi_{\beta}\right\rangle
   \left | \overline{\psi}_{\beta}\right\rangle.
\end{equation}
Using the new basis in Eq.\ (\ref{eq:newbasis1}), we see that
Eq.\ (\ref{eq:newnorm}) allows us to define another
orthogonal basis $h$
\begin{equation}
  h_{\alpha}^{\lambda}=\sqrt{\mu_{\alpha}^2+1}\sum_{\beta}
  d_{\beta}^{\alpha}b_{\beta}^{\lambda}\frac{1}{\sqrt{N_{\lambda}}}
  =\frac{1}{\sqrt{\mu_{\alpha}^2+1}}\sum_{\beta}
  d_{\beta}^{\alpha}\overline{b}_{\beta}^{\lambda}\sqrt{N_{\lambda}},
  \label{eq:hbasis}
\end{equation}
where we have used the orthogonality properties of the vectors
involved. The vector $h$ was used by the authors of Ref.\ \cite{kehlsok93}
to obtain a hermitian effective interaction as
\begin{equation}
   \left\langle \psi_{\alpha}\right |
   V_{\mathrm{eff}}^{\mathrm{(her)}}\left | \psi_{\beta}\right\rangle=
   \frac{\sqrt{\mu_{\alpha}^2+1}\left\langle \psi_{\alpha} \right |
   V_{\mathrm{eff}}^{\mathrm{(nher)}}\left | \psi_{\beta}\right\rangle
   +\sqrt{\mu_{\beta}^2+1}\left\langle \psi_{\alpha}\right |
   V_{\mathrm{eff}}^{\dagger\mathrm{(nher)}}\left |
   \psi_{\beta} \right\rangle}
   {\sqrt{\mu_{\alpha}^2+1}+\sqrt{\mu_{\beta}^2+1} },
   \label{eq:hermitian}
\end{equation}
where (her) and (nher) stand for hermitian and non-hermitian
respectively. 
This equation is rather simple to compute, since
we can use the folded-diagram method
to obtain the non-hermitian part.
To obtain the total hermitian effective interaction, we have to
add the $H_0$ term. The above equation is manifestly hermitian.
Other discussion of the hermiticity problem can be found
in Refs.\ \cite{lm85,arponen97}.
The remaining question is how to evaluate the $\hat{Q}$-box. 
Obviously, we are not in the position where we can evaluate it
exactly, i.e., to include all possible many-body terms.
Rather, we have to truncate somewhere. 
Several possible approaches exist, but all have in common that
there is no clear way which tells us where to stop.
However, as argued by the authors of Ref.\ \cite{jls82}, there is a
minimal class of diagrams which need to be included 
in order to fulfull necessary conditions. This class
of diagrams includes both diagrams which account for short-range
correlations such as the $G$-matrix and long-range
correlations such as those accounted for by various 
core-polarization terms. 
The importance of such diagrams has been extensively documented
in the literature and examples can be found
in Refs.\ \cite{hko95,eo77}. In Ref.\ \cite{hko95} we included
all core-polarization contributions to third-order
in the $G$-matrix, in addition to including other diagrams
which account for short-range correlations as well. 

In the next subsection we present results from recent 
large scale shell-model Monte Carlo calculations \cite{drhklz99}
for light nuclei in the $1s0d$ and $1p0f$ shells based 
on an effective interaction for the two shells.
The effective interaction was derived following the methods
outlined hitherto in this section, taking into account 
the non-hermiticity which arises when dealing 
with more than one major shell.
The closed shell core is $^{16}$O and all diagrams through
second-order where employed in the definition of the
$\hat{Q}$-box. Folded diagrams were calculated using the abovementioned
folded-diagrams method and the interaction was made explicitely
hermitian through Eq.\ (\ref{eq:hermitian}). A discussion of results
for Sn isotopes will also be given.

\subsection{Selected applications}

\subsubsection{Neutron-rich nuclei in the $1s0d$-$1p0f$ shells}

Studies of extremely neutron-rich nuclei have revealed a number of
intriguing new phenomena.  Two sets of these nuclei that have received
particular attention are those with neutron number $N$ in the vicinity
of the $1s0d$ and $0f_{7/2}$ shell closures ($N \approx 20$ and $N
\approx 28$).  Experimental studies of neutron-rich Mg and Na isotopes
indicate the onset of deformation, as well as the modification of the
$N = 20$ shell gap for $^{32}$Mg and nearby nuclei \cite{r:motobayashi}.
Inspired by the rich set of phenomena occurring near the $N = 20$
shell closure when $N \gg Z$, attention has been directed to nuclei
near the $N = 28$ (sub)shell closure for a number of S and Ar isotopes
\cite{r:brown1,r:brown2} where similar, but less dramatic, effects
have been seen as well.
\begin{table}[hbtp]
\begin{center}
\caption{The computed and measured values of $B(E2)$ for
the nuclei in this study using $e_p=1.5$ and $e_n=0.5$.
}
\begin{tabular}{|cccc|}\hline
 & $B(E2; 0^+_{gs} \rightarrow 2^+_1)_{Expt}$ & $B(E2, total)_{SMMC}$ &
  $B(E2; 0^+_{gs} \rightarrow 2^+_1)$  \\\hline
 $^{22}$Mg & $458 \pm 183$ & $334 \pm 27 $
    & \\
 $^{30}$Ne & & $303 \pm 32$
    & 342 \cite{r:fukunishi},171 \cite{r:poves2}  \\
 $^{32}$Mg & $454 \pm 78$ \cite{r:motobayashi} & $494 \pm 44 $
   & 448 \cite{r:fukunishi},205 \cite{r:poves2} \\
 $^{36}$Ar & $296.56 \pm 28.3$ \cite{r:ensdf} & $174 \pm 48$
    & \\
 $^{40}$S & $334 \pm 36$ \cite{r:brown1} & $270 \pm 66$
    & 398 \cite{r:brown2},390 \cite{r:retamosa}  \\
 $^{42}$S & $397 \pm 63$ \cite{r:brown1} & $194 \pm 64$
   & 372 \cite{r:brown2},465 \cite{r:retamosa}  \\
 $^{42}$Si &  &  $445 \pm 62$
    & 260 \cite{r:retamosa}  \\
 $^{44}$S & $314 \pm 88$ \cite{r:brown2} & $274 \pm 68$
    & 271 \cite{r:brown2},390 \cite{r:retamosa}  \\
 $^{44}$Ti & $610 \pm 150$ \cite{r:raman} & $692 \pm 63$
    &  \\
 $^{46}$Ar & $196 \pm 39$ \cite{r:brown1} & $369 \pm 77 $
    & 460 \cite{r:brown1},455 \cite{r:retamosa}  \\\hline
\end{tabular}
\end{center}
\label{t:tab1}
\end{table}
In parallel with the experimental efforts, there have been several
theoretical studies seeking to understand and, in some cases, predict
properties of these unstable nuclei.  Both mean-field
\cite{r:werner,r:campi} and shell-model calculations
\cite{r:brown1,r:brown2,r:wbmb,r:poves1,r:fukunishi,r:retamosa,r:caurier}
have been proposed. The latter require a severe truncation to
achieve tractable model spaces, since the successful description of these
nuclei involves active nucleons in both the $1s0d$- and the $1p0f$-shells.
The natural basis for the problem is therefore the full $1s0d$-$1p0f$
space, which puts it out of reach of exact diagonalization on current
hardware\footnote{For a treatment of the c.m. problem, see Ref.\ 
\cite{drhklz99}.}.

Shell-Model Monte Carlo (SMMC) methods
\cite{r:david97,r:smmc_ar,r:lang} offer an alternative to direct
diagonalization when the bases become very large. Though SMMC provides
limited detailed spectroscopic information, it can predict, with good
accuracy, overall nuclear properties such as masses, total strengths,
strength distributions, and deformation, precisely those quantities
probed by the recent experiments.

There is limited experimental information about the highly unstable,
neutron-rich nuclei under consideration.  In many cases only the mass,
excitation energy of the first excited state, the $B(E2)$ to that state,
and the $\beta$-decay rate is known, and not even all of this
information is available in some cases.  From the
measured $B(E2)$, an estimate of the nuclear deformation parameter,
$\beta_2$, has been obtained via the usual relation
\begin{equation}
\beta_2 = 4 \pi \sqrt{B(E2; 0^+_{gs} \rightarrow 2^+_1)}/3 Z R_0^2 e
\end{equation}
with $R_0 = 1.2 A^{1/3}$ fm and $B(E2)$ given in $e^2$fm$^4$.

Much of the interest in the region stems from the unexpectedly large
values of the deduced $\beta_2$, results which suggest the onset of
deformation and have led to speculations about the vanishing of the $N
= 20$ and $N = 28$ shell gaps.  The lowering in energy of the 2$^+_1$
state supports this interpretation.  The most thoroughly studied case,
and the one which most convincingly demonstrates these phenomena, is
$^{32}$Mg with its extremely large $B(E2) = 454 \pm 78 \, e^2$fm$^4$ and
corresponding $\beta_2 = 0.513$ \cite{r:motobayashi}; however, a word of
caution is necessary when deciding on the basis of this
limited information that we are in the presence of well-deformed
rotors: for $^{22}$Mg, we would obtain $\beta_2 = 0.67$, even more
spectacular, and for $^{12}$C, $\beta_2 = 0.8$, well above the
superdeformed bands.

Most of the measured observables can be calculated within the SMMC
framework.  It is well known that in {\it deformed} nuclei the total
$B(E2)$ strength is almost saturated by the $0^+_{gs} \rightarrow
2_1^+$ transition (typically 80\% to 90\% of the strength lies in this
transition).  
Thus the total strength calculated by SMMC should only
slightly overestimate the strength of the measured transition.  In
Table 1 the SMMC computed values of $B(E2, total)$ are
compared both to the experimental $B(E2; 0^+_{gs} \rightarrow 2^+_1)$
values and to the values found in various truncated shell-model
calculations.  Reasonable agreement with experimental data across the
space is obtained when one chooses effective charges of $e_p=1.5$ and
$e_n=0.5$. 
All of the theoretical
calculations require excitations to the $1p0f$-shell before reasonable
values can be obtained.  We note a general agreement among all
calculations of the $B(E2)$ for $^{46}$Ar, although they are typically
larger than experimental data would suggest. We also note a somewhat
lower value of the $B(E2)$ in this calculation as compared to
experiment and other theoretical calculations in the case of $^{42}$S.
Table 2 gives selected occupation numbers for the nuclei
considered.  We first note a difficulty in extrapolating some of the
occupations where the number of particles is nearly zero.  This leads
to a systematic error bar that we estimate at $\pm 0.2$ for all
occupations shown, while the statistical error bar is quoted in the
table. The extrapolations for occupation numbers were principally
linear. Table 2 shows that $^{22}$Mg remains as an almost
pure $sd$-shell nucleus, as expected.  We also see that the protons in
$^{30}$Ne, $^{32}$Mg, and $^{42}$Si are almost entirely confined to the
$sd$ shell.  This latter is a pleasing result in at least two regards.
First, it shows that the interaction does not mix the two shells to an
unrealistically large extent.  Second, if spurious c.m.\ contamination
were a severe problem, we would expect to see a larger proton
$0f_{7/2}$ population for these nuclei due to the $0d_{5/2}$-$0f_{7/2}$
``transition'' mediated by the center-of-mass creation operator.  The fact that
there is little proton $f_{7/2}$ occupation for these nuclei confirms
that the c.m.\ contamination is under reasonable control.
See Ref.\ \cite{drhklz99} for further details.
\begin{table}[hbtp]
\begin{center}
\caption{The calculated SMMC neutron and
proton occupation numbers for the $sd$ shell, the
$0f_{7/2}$ sub-shell, and the remaining orbitals of the
$pf$ shell.  The statistical errors are given for linear
extrapolations. A systematic error of $\pm 0.2$ should also
be included. The first row represents neutron results, while the 
second row represents protons.}
\begin{tabular}{|ccccc|}\hline
 & $N,Z$ & $1s0d$ & $0f_{7/2}$ & $1p0f_{5/2}$\\ \hline
$^{22}$Mg & 10,12 & $3.93 \pm 0.02$ & $0.1 \pm  0.02$ &
  $-0.05 \pm 0.01$ \\ &&  $2.04 \pm 0.02$ & $0.00 \pm 0.01$ &
  $-0.05 \pm 0.01$ \\
$^{30}$Ne & 20,10 & $9.95 \pm 0.03$ & $2.32 \pm 0.03$ &
  $-0.26 \pm 0.02$ \\ && $2.03 \pm 0.02$ & $-0.01 \pm 0.01$ &
  $-0.02 \pm 0.01$ \\
$^{32}$Mg & 20,12 & $9.84 \pm 0.03$ & $ 2.37 \pm 0.03$ &
  $-0.21 \pm 0.02$ \\ && $3.99 \pm 0.03$ & $0.05 \pm 0.02$ &
  $-0.05 \pm 0.01$ \\
$^{36}$Ar & 18,18 & $9.07 \pm 0.03$ & $1.08 \pm 0.02$ &
  $-0.15 \pm 0.02$ \\ && $9.07 \pm 0.03$ & $1.08 \pm 0.02$ &
  $-0.15 \pm 0.02$ \\
$^{40}$S & 24,16 & $11.00 \pm 0.03$ & $ 5.00 \pm 0.03 $ &
  $-0.01\pm 0.02$ \\ && $7.57 \pm 0.04$ & $0.54 \pm 0.02$ &
  $-0.12 \pm 0.02$ \\
$^{42}$Si & 28,14 & $11.77 \pm 0.02$ & $7.34 \pm 0.02$ &
  $0.90 \pm 0.03$ \\ && $5.79 \pm 0.03$ & $0.25 \pm 0.02$ &
  $-0.07 \pm 0.01$ \\
$^{42}$S & 26,16 & $11.41 \pm 0.02$ & $6.33 \pm 0.02$ &
  $0.25 \pm 0.03$ \\ && $7.49 \pm 0.03$ & $0.58 \pm 0.02$ &
  $-0.09 \pm 0.02$ \\
$^{44}$S & 28,16 & $11.74 \pm 0.02$ & $7.18 \pm 0.02$ &
  $1.06 \pm 0.03$ \\ && $7.54 \pm 0.03$ & $0.56 \pm 0.02$ &
  $-0.12 \pm 0.02$ \\
$^{44}$Ti & 22,22 & $10.42 \pm 0.03$ & $3.58 \pm 0.02$ &
  $0.00 \pm 0.02$\\ & & $10.42 \pm 0.03$ & $3.58 \pm 0.02$ &
  $0.00 \pm 0.02$ \\
$^{46}$Ar & 28,18 & $11.64 \pm 0.02$ & $7.13 \pm 0.02$ &
  $1.23 \pm 0.03$ \\ && $8.74 \pm 0.03$ & $1.34 \pm 0.02$ &
  $-0.08 \pm 0.02$ \\\hline
\end{tabular}
\end{center}
\label{t:tab3}
\end{table}
An interesting feature of Table 2 lies in the neutron
occupations of the $N = 20$ nuclei ($^{30}$Ne and $^{32}$Mg) and the $N =
28$ nuclei ($^{42}$Si, $^{44}$S, and $^{46}$Ar).  The neutron
occupations of the two $N = 20$ nuclei are quite similar, confirming
the finding of Fukunishi {\it et al.} \cite{r:fukunishi} and Poves and
Retamosa \cite{r:poves1} that the $N= 20$ shell gap is modified.  In
fact, the neutron $0f_{7/2}$ orbital contains approximately two
particles before the $N=20$ closure, thus behaving like an intruder
single-particle state.  Furthermore, we see that 2p-2h excitations
dominate although higher excitations also play some role.  We also see
that the neutrons occupying the $1p0f$-shell in $N=20$ systems are
principally confined to the $0f_{7/2}$ sub-shell.
The conclusions that follow from looking at nuclei with $N > 20$,
particularly those with $N = 28$, are that the $N = 20$ shell is nearly
completely closed at this point, and that the $N=28$ closure shell is
reasonably robust, although approximately one neutron occupies the upper
part of the $1p0f$ shell. Coupling of the protons with the low-lying
neutron excitations probably accounts for the relatively large
$B(E2)$, without the need of invoking rotational behavior.
\begin{table}[hbtp]
\begin{center}
\caption{The calculated total Gamow-Teller strength, $GT^-$,
from this study.  The results of other studies, when
available, are presented for comparison.
}
\begin{tabular}{|ccc|}\hline
 Nucleus & SMMC & Other \\
\hline
 $^{22}$Mg & $0.578 \pm  0.06$  & \\
 $^{30}$Ne & $29.41 \pm 0.25$ & \\
 $^{32}$Mg & $24.00 \pm 0.34$ & \\
 $^{36}$Ar & $2.13 \pm  0.61$ & \\
 $^{40}$S  & $22.19 \pm 0.44$ & 22.87 \cite{r:retamosa} \\
 $^{42}$S  & $28.13 \pm 0.42$ & 28.89 \cite{r:retamosa} \\
 $^{42}$Si & $40.61 \pm 0.34$ & \\
 $^{44}$S  & $34.59 \pm 0.39$ & 34.93 \cite{r:retamosa} \\
 $^{44}$Ti & $4.64 \pm  0.66$ & \\
 $^{46}$Ar & $29.07 \pm 0.44$ & 28.84 \cite{r:retamosa} \\\hline
\end{tabular}
\end{center}
\label{t:tab4}
\end{table}
In Table 3 we show the SMMC total Gamow-Teller (GT$^-$)
strength.  We compare our results to those of previous truncated
calculations, where available.  In all cases, our results are slightly
smaller than, but in good accord with, other calculations.  Since we
do not calculate the strength function, we do not compute
$\beta$-decay lifetimes.

\subsubsection{Heavy Sn isotopes}

Effective two-hole matrix elements are calculated based on 
a $Z = 50, \quad N = 82$ asymmetric core and 
with the active $P$-space for holes
based on the $2s_{1/2}$, $1d_{5/2}$, $1d_{3/2}$, $0g_{7/2}$ and $0h_{11/2}$
hole orbits. The CD-Bonn model for the NN interaction was employed
\cite{cdbonn} and all diagrams through third-order in the $G$-matrix
were included in the evaluation of the $\hat{Q}$-box. Folded diagrams
were again included through the method exposed above, see
Refs.\ \cite{hko95,ehho98} for further details.
The corresponding single-hole energies
$\varepsilon(d_{3/2}^{+}) = 0.00$~MeV, 
 $\varepsilon(h_{11/2}^{-}) = 0.242$~MeV, 
$\varepsilon(s_{1/2}^{+}) = 0.332$~MeV,
$\varepsilon(d_{5/2}^{+}) = 1.655$~MeV and  
$\varepsilon(g_{7/2}^{+}) = 2.434$~MeV
are taken from \mbox{Ref. \cite{jan}}
and the shell model calculation amounts to studying
valence neutron holes outside this core.
The shell model problem requires the solution of a real symmetric
$n \times n$ matrix eigenvalue equation
\begin{equation}
       \widetilde{H}\left | \Psi_k\right\rangle  = 
       E_k \left | \Psi_k\right\rangle .
       \label{eq:shell_model}
\end{equation}
where for the present cases the dimension of 
the $P$-space reaches $n \approx 2 \times 10^{7}$.
At present our basic approach in finding 
solutions to Eq.\ (\ref{eq:shell_model})
is the Lanczos algorithm; an iterative method
which gives the solution of the lowest eigenstates. This method was 
already applied to nuclear physics problems by Whitehead et al. 
in 1977. The technique is described in detail in Ref.\ \cite{whit77},
see also Ref.\ \cite{ehho95}.
The results of the shell model calculation are presented in Table 4.
All experimental information in the present analysis is taken 
from the data base of the 
National Nuclear Data Center at \mbox{Brookhaven \cite{brook}}. 
\begin{table}[t]
\caption{Exitation spectra for the heavy Sn isotopes. \label{tab:2}}
\vspace{0.2cm}
\begin{center}
\footnotesize
\begin{tabular}{|cccc|cccc|}
\hline
&&&&&&&\\[-5pt]
\multicolumn{4}{|c|}{$^{130}$Sn}&\multicolumn{4}{|c|}{$^{128}$Sn}\\
$J^{\pi}$&Exp.&$J^{\pi}$&Theory&$J^{\pi}$&Exp.&$J^{\pi}$&Theory\\
\hline
&&&&&&&\\[-3pt]
$(2^{+})$ & $1.22$ & $2^{+}$   & $1.46$ &$(2^{+})$ & $1.17$ & $2^{+}$ & $1.28$\\
$(4^{+})$ & $2.00$ & $4^{+}$   & $2.39$ &$(4^{+})$ & $2.00$ & $4^{+}$ & $2.18$\\
$(6^{+})$ & $2.26$ & $6^{+}$   & $2.64$ &$(6^{+})$ & $2.38$ & $6^{+}$ & $2.53$\\[3pt]
\hline \hline
&&&&&&&\\[-5pt]
\multicolumn{4}{|c|}{$^{126}$Sn}&\multicolumn{4}{|c|}{$^{124}$Sn}\\
$J^{\pi}$&Exp.&$J^{\pi}$&Theory&$J^{\pi}$&Exp.&$J^{\pi}$&Theory\\
\hline
&&&&&&&\\[-3pt]
$2^{+}$ & $1.14$ & $2^{+}$   & $1.21$ &$2^{+}$ & $1.13$ & $2^{+}$ & $1.17$\\
$4^{+}$ & $2.05$ & $4^{+}$   & $2.21$ &$4^{+}$ & $2.10$ & $4^{+}$ & $2.26$\\
$     $ &        & $6^{+}$   & $2.61$ &        &        & $6^{+}$ & $2.70$\\[3pt]
\hline \hline
&&&&&&&\\[-5pt]
\multicolumn{4}{|c|}{$^{122}$Sn}&\multicolumn{4}{|c|}{$^{120}$Sn}\\
$J^{\pi}$&Exp.&$J^{\pi}$&Theory&$J^{\pi}$&Exp.&$J^{\pi}$&Theory\\
\hline
&&&&&&&\\[-3pt]
$2^{+}$   & $1.14$ & $2^{+}$   & $1.15$ & $2^{+}$  & $1.17$ & $2^{+}$ & $1.14$\\
$4^{+}$   & $2.14$ & $4^{+}$   & $2.30$ & $4^{+}$  & $2.19$ & $4^{+}$ & $2.30$\\
$6^{+}$   & $2.56$ & $6^{+}$   & $2.78$ &          &        & $6^{+}$ & $2.86$\\[3pt]
\hline \hline
&&&&&&&\\[-5pt]
\multicolumn{4}{|c|}{$^{118}$Sn}&\multicolumn{4}{|c|}{$^{116}$Sn}\\
$J^{\pi}$&Exp.&$J^{\pi}$&Theory&$J^{\pi}$&Exp.&$J^{\pi}$&Theory\\
\hline
&&&&&&&\\[-3pt]
$2^{+}$   & $1.22$ & $2^{+}$   & $1.15$ & $2^{+}$  & $1.30$ & $2^{+}$ & $1.17$\\[3pt]\hline
\end{tabular}
\end{center}
\end{table}

The isotopes above $^{116}$Sn (heavy Sn) are treated based 
on the asymmetric $Z = 50, N = 82$ core. 
This simplifies the shell model calculation,
but in addition it is of interest to see how successful a hole-hole effective 
interaction calculated with respect to $^{132}$Sn  is.

Only some selected states are displayed. First of all, the well-known 
near constant $0^{+} - 2^{+}$ spacing is well reproduced. 
all the way down to 
$^{116}$Sn.
Also the additional calculated states are in very good agreement
with experiment. However more detailed analysis of 
the results close to $^{116}$Sn
indicates that our effective two-particle interaction 
has difficulties in reproducing
the shell closure which is believed to occur in this region. 
The increase of the 
the $0^{+} - 2^{+}$ splitting is not as sharp as found 
experimentally, even if the 
phenomenon is rather weak in the case of Sn.
We have observed a similar feature around $^{48}$Ca \cite{hko95}
which is generally agreed
to be a good closed shell nucleus. There the deviation 
between theory and experiment 
is severe. Preliminary analysis indicates that our effective interaction
may be slightly too actractive when the two particles 
occupy different single-particle orbits.
This may be related to the radial wave functions which in our calculation are 
chosen to be harmonic oscillator functions.

\subsection{Inclusion of hole-hole contributions and 
single-particle propagators}

With the $G$-matrix defined according 
to the double-partitioned scheme we can easily solve Eq.\ (\ref{eq:first12})
through matrix inversion. The number of hole-hole and particle-particle
configurations is then rather small, typically smaller than
$\sim  100$, and a matrix inversion
is then rather trivial. 
Before we discuss the solution of Eq.\ (\ref{eq:first12}), it is always
instructive to consider the contributions to second order in 
perturbation theory, i.e., diagrams (a) and (b) of Fig.\
\ref{fig:gamma12}. The external legs can be 
particle states or hole states. Diagram (a) reads
\begin{equation}
      (a)=\frac{1}{2}\sum_{pq}V^{[12]}_{12pq J}
      \frac{1}{s-\varepsilon_p-
                \varepsilon_q} V^{[12]}_{pq34 J},
      \label{eq:secondg}
\end{equation}
and 
\begin{equation}
      (b)=\frac{1}{2}\sum_{\alpha\beta}V^{[12]}_{12\alpha\beta J}
      \frac{1}{-s+\varepsilon_{\alpha}+
                \varepsilon_{\beta}} V^{[12]}_{\alpha\beta 34 J}.
\end{equation}
We note here the minus sign in the energy denominator,
since in the latter expression we are using
the hole-hole term of the propagator of
Eq.\ (\ref{eq:paulioperator12}).
If we use a double-partitioned $G$-matrix for say $^{16}$O and
are interested in an effective valence space interaction
for the $1s0d$-shell\footnote{This means that the labels $1234$ will
refer to particle states in the $1s0d$-shell.}, then typically
the single-particle orbits of the intermediate states will be represented
by states in the $1p0f$ major shell. Hole states are then defined
by single-particle states in the $0s$ and $0p$ shells.
Clearly, the number of two-body intermediate states is rather limited. 
To third order we have diagrams like (c) and (d) of Fig.\
\ref{fig:gamma12}. Diagram (c) is just the third-order equivalent of
Eq.\ (\ref{eq:secondg}) and reads
\begin{equation}
      (c)=\frac{1}{4}\sum_{pqrw}V^{[12]}_{12pq J}
      \frac{1}{s-\varepsilon_p-
                \varepsilon_q} 
      V^{[12]}_{pqrw J}
      \frac{1}{s-\varepsilon_r-
                \varepsilon_w} 
       V^{[12]}_{rw34 J},
      \label{eq:thirdg}
\end{equation}
while diagram (d) contains both a two-particle and a two-hole 
intermediate state and reads
\begin{equation}
      (d)=\frac{1}{4}\sum_{\alpha\beta pq}V^{[12]}_{12pq J}
      \frac{1}{s-\varepsilon_p-
                \varepsilon_q+\varepsilon_{\alpha}+
                \varepsilon_{\beta}} 
      V^{[12]}_{pq\alpha\beta J}
      \frac{1}{s-\varepsilon_p-
                \varepsilon_q} 
       V^{[12]}_{\alpha\beta 34 J}.
      \label{eq:thirdg2h}
\end{equation}
Thus, solving Eq.\ (\ref{eq:first12}) will then
yield contributions to the effective interaction
such as the above expressions. 
 
Here we have also tacitly assumed that the energy denominators
do not diverge, i.e., we have chosen an energy $s$
so that we avoid the poles. This has always been the standard
approach in calculations of shell-model 
effective interactions. To give an example, consider
now diagram (b) and suppose that we are using harmonic
oscillator wave functions. Let us also assume that the two
hole states are from the $0p$-shell and that the valence
particles are in the $1s0d$-shell. If we rescale the
energies of the valence space to zero, then the two-hole
state would yield $-28$ MeV with an oscillator parameter
$b=1.72$ fm. If $s=-28$, the denominator diverges.
In this case it is rather easy to obtain the imaginary
part, and even if we were to chose $s$ different
from $-28$ MeV, the imaginary part will influence the real
part of the effective interaction through dispersion
relations, see e.g., Refs.\ \cite{angels88,rpd89,ms92}.
It is therefore at best just a first approximation
to neglect the imaginary term.
Moreover, if we solve Dyson's equation for the 
self-energy, the single-particle energies may contain
an imaginary part. 
Technically it is however not difficult to deal
with imaginary contributions, one needs to 
invert a complex  matrix rather than a real one.
However, care must be exercised in localizing poles,
see e.g., Ref.\ \cite{landau97} for a
computational approach to this problem. In our actual
calculations we will also follow Ref.\
\cite{landau97}. These technicalities
will however be described elsewhere \cite{mhj99}. 

Using the double-partitioned $G$-matrix, we can then rewrite
Eq.\ (\ref{eq:first12}) as
\begin{equation}
      \Gamma^{[12]}_{1234J}(s) = 
      G^{[12]}_{1234J}+\frac{1}{2}
      \sum_{56}
      G^{[12]}_{1256J}\hat{{\cal G}}^{[12]}
      \Gamma^{[12]}_{5634J}(s),
      \label{eq:newfirst12}
\end{equation}
where $G^{[12]}$ is just the double-partitioned $G$-matrix
discussed above. It is also energy dependent,
in contrast to $V$.
In case we were to employ this equation for effective
interactions in the $1s0d$-shell, the intermediate two-particle
states would then come from just e.g., the $1p0f$-shell.
This equation, which now is solved within a much smaller space
than the original one spanned by the total $Q_{pp}$, allows
clearly for computationally amenable solutions. It corresponds
to the so-called {\em model-space approach} to the solution
of the Feynman-Galitskii equations as advocated by
Kuo and co-workers, see e.g., Ref.\ \cite{kt94}
for more details. Thus, a possible approach would consist
of the following steps 
\begin{enumerate}
\item Solve the $G$-matrix equation from Eq.\ (\ref{eq:gmod})
      using the double-partitioning scheme.
\item The next step is then to solve Eq.\ (\ref{eq:first12})
      and Dyson's equation for the self-energy.
\item This scheme is iterated till self-consistency is achieved,
      see the discussion below. 
\end{enumerate}

We will however not employ this {\em model-space} scheme in our
actual calculations. There are several reasons for not doing so.

Let us first assume that we omit the $[13]$ and $[14]$ channels in our
iterative scheme for Eq.\ (\ref{eq:newfirst12}). 
The next iteration of Eq.\ (\ref{eq:newfirst12}) 
would then look like 
\begin{equation}
      \Gamma^{[12]}_{(1)} = 
      \Gamma_{(0)}^{[12]}+
      \Gamma_{(0)}^{[12]}
       \hat{{\cal G}}^{[12]}\Gamma^{[12]}_{(1)},
      \label{eq:second12}
\end{equation}
where the vertex function $\Gamma_{(0)}^{[12]}$ is the solution
of Eq.\ (\ref{eq:newfirst12}). However, we cannot define
the ``bare'' vertex $\Gamma_{(0)}^{[12]}$ to be the 
solution  of Eq.\ (\ref{eq:newfirst12}) simply because then we 
would be double-counting contributions.
Thus, $\Gamma_{(0)}^{[12]}$ has to equal the $G$-matrix.
The only change in Eq.\ (\ref{eq:second12}) 
arises from the solution of Dyson's equation
and thereby new single-particle energies. 

Let us then for the sake of simplicity assume that the 
single-particle energies are just the Hartree-Fock
solutions. The problem we are aiming at arises at the 
Hartree-Fock level.
In order to obtain Hartree-Fock solutions which
are independent of the chosen harmonic oscillator
parameter $b$, we typically need to include single-particle
orbits from quite many major shells.
Typical constraints we have found when we do so-called
Brueckner-Hartree-Fock (BHF) calculations for finite nuclei is that
we need at least $2n+l \leq 20$ in order to obtain
a result which is independent of the chosen 
$b$\footnote{Throughout this work our unperturbed single-particle
basis par excellence will always be that of the harmonic
oscillator.} value. The way we solve the  
BHF equations is to expand the new single-particle
wave functions $\psi_{\lambda}$, with $\lambda$ 
representing the quantum numbers $nlj$,
in terms of harmonic oscillator wave functions,
i.e.,
\begin{equation}
     \left | \psi_{\lambda}\right\rangle=
     \sum_{\alpha =1}^{2n+l\leq 20} 
     C_{\alpha}^{(\lambda )}\left | \phi_{\alpha}\right\rangle
     \label{eq:selfconstbasis}
\end{equation}
where $\phi_{\alpha}$ are the harmonic oscillator wave functions
with quantum numbers $\alpha=nlj$ and $C$ are the coefficients
to be varied in the Hartree-Fock calculations.
The single-particle energies at the 
Hartree-Fock level are just
\begin{equation}
  \varepsilon_{\alpha}=t_{\alpha}+
   \sum_h \bra{\alpha h} G(\varepsilon_{\alpha}+
                           \varepsilon_h)
          \ket{\alpha h},
\end{equation}
where the single-particle states are just those
of the harmonic oscillator. The $G$-matrix used in the 
first iteration in the BHF calculation is the one given
by the solution of Eq.\ (\ref{eq:gmod}).
The coefficients 
$C_{\alpha}$ can then be obtained by diagonalizing
a matrix of dimension $N\times N$, where $N$ is the number
of single-particle orbits with the same $lj$ values.
As an example, suppose that we are considering the 
$s_{1/2}$ state with $l=0$ and $j=1/2$. With the above
requirement $2n+l\leq 20$, we may have
that $N_{\mathrm{max}}=10$, the dimensionality being equal to 
the quantum number $n$. 
The way to proceed in a BHF calculation is to calculate
the reference $G$-matrix $G_F$ in Eq.\ (\ref{eq:freeg})
once and for all\footnote{This matrix is typically
set up in the relative and c.m. system and calculated
only once and for all 
for $b=1$ fm. Other $b$-values involve simply
a multiplication with a constant. The matrix $G_F$ can therefore
be used for other mass areas as well.} .
Thereafter, the change in single-particle wave functions
is introduced in the calculations of Eq.\ (\ref{eq:gmod}).

We see then that if we choose to do 
the Hartree-Fock self-consistency
employing the double-partitioned $G$-matrix and summing
the pphh diagrams as in Eq.\ (\ref{eq:second12}), 
our single-particle basis will just be
defined by the $0s$, $0p$, $1s0d$ and $1p0f$ shells in the case
of an effective interaction in the $1s0d$-shell. 
{\em This is simply
not enough in order to obtain a stable Brueckner-Hartree-Fock
result.}  

The reader could infer that why do we not perform first
a BHF calculation for the $G$-matrix in Eq.\ (\ref{eq:gmod}),
and then solve Eq.\ (\ref{eq:first12}) and stop there.
This would be in line with the abovementioned 
{\em model-space approach} of Kuo and co-workers \cite{kt94}.
However, if one calculates the self-energy by only
including particle-particle intermediate states
in the vertex function $\Gamma$, which is the case
if we do a standard BHF calculation, one may seriously violate 
various sum rules, as demonstrated in Refs.\ 
\cite{ms92,mahaux85}. 
Thus, to respect sum rules such as the conservation
of number of particles constrains severely the way we
solve Eqs.\ (\ref{eq:first12}) and (\ref{eq:dyson12}).
The vertex function needs both particle-particle and hole-hole
intermediate states in order to eventually
satisfy e.g., the conservation
of the number of particles.

Moreover, our interest lies in solving
the Parquet equations. This
entails simply that we perform the above
self-consistency. 
It is therefore not only a matter of many-body aestethics for
embarking on the solution of the Parquet equations.
We are also able to satisfy various sum rules.
In section \ref{sec:sec4} we will also show that
at every level of approximation, the solution
of the equations in the $[13]$ and $[14]$ channels will
also result in an antisymmetric vertex function in these
channels. 

In section \ref{sec:sec5} we will come back to the
technical solution of Eq.\ (\ref{eq:first12}) and its
iterations.

\section{Screening corrections and vertex renormalization, the equations
for the $[13]$ and  $[14]$ channels}
\label{sec:sec4}

We start as in the previous section with the definition of the interaction
vertices in the $[13]$ and $[14]$ channels and the corresponding
integral equations. Thereafter, we discuss various approximations
to these equations such as the summation of TDA and RPA diagrams.
Eventually, the aim is to merge the discussion in this section and 
the preceeding one into equations for a self-consistent scheme which combines 
all three channels, namely the so-called set of Parquet equations to
be discussed in section \ref{sec:sec5}. 

\begin{figure}[hbtp]
      \setlength{\unitlength}{1mm}
      \begin{picture}(100,20)
      \put(35,0){\epsfxsize=7cm \epsfbox{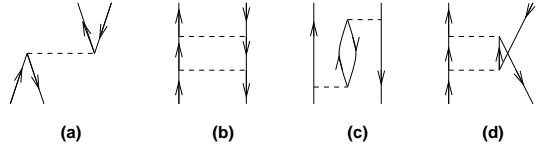}}
      \end{picture}
      \caption{Examples of irreducible and reducible diagrams in the 
               $[13]$ and $[14]$ channels. See text for further details.}
      \label{fig:1314channel}
\end{figure}

The equations for the renormalized vertex in the $[13]$ and $[14]$
channels have the same form as Eq.\ (\ref{eq:schematic12}), namely
\begin{equation}
     \Gamma^{[13]}=V^{[13]}+V^{[13]}(gg)\Gamma^{[13]},
\end{equation}
and 
\begin{equation}
     \Gamma^{[14]}=V^{[14]}+V^{[14]}(gg)\Gamma^{[14]}.
\end{equation}
The matrix elements which enter are however
defined differently and the irreducible
diagrams of $V^{[13]}$ and $V^{[14]}$ can obviously not be the same.
With irreducible in the $[13]$ channel we will mean a diagram, which by
cutting an internal particle-hole pair, cannot be separated into a piece
containing the external legs $1,3$ and another piece containing
$2,4$ as external legs. The definition for the irreducible vertex in the 
$[14]$ channel is similar and we illustrate these differences in Fig.\
\ref{fig:1314channel}. 
Diagram (a) is just the lowest-order
interaction in the $[13]$ channel and is therefore irreducible.
Diagram (b) is an irreducible diagram in the $[13]$ channel,
whereas it is reducible in the $[14]$ channel. Diagram (c) is in turn
irreducible in the $[14]$ channel and reducible in the $[13]$ channel. 
Diagram (d) is an example of a diagram which is irreducible in both
channels. This diagram stems from the $[12]$ channel.

\begin{figure}[hbtp]
      \setlength{\unitlength}{1mm}
      \begin{picture}(100,70)
      \put(35,0){\epsfxsize=7cm \epsfbox{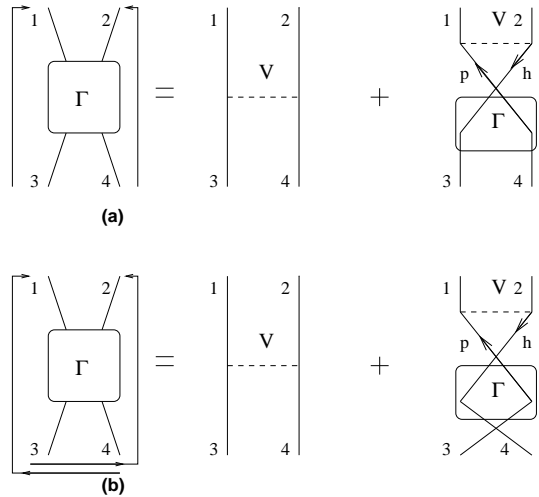}}
      \end{picture}
      \caption{(a) shows the structure of the integral equation for 
               the interaction vertex in the $[13]$ channel. (b) represents
               the integral channel for the $[14]$ channel.
               The coupling order is displayed as well.}
      \label{fig:figs1314}
\end{figure}

The energy variables in these channels are, following
Fig.\ \ref{fig:channelsdef} and Eqs.\ (\ref{eq:13channel}) and
(\ref{eq:14channel}),
\begin{equation} 
     t=\varepsilon_3-\varepsilon_1=\varepsilon_2-\varepsilon_4,
\end{equation}    
for the $[13]$ channel and
\begin{equation}
     u=\varepsilon_1-\varepsilon_4=\varepsilon_3-\varepsilon_2,
\end{equation}
for the $[14]$ channel.
Defining the unperturbed particle-hole propagators 
in the energy representation as \cite{kt94}
\begin{equation}
    \hat{{\cal G}}^{[13]}=
    \frac{Q^{[13]}_{\mathrm{ph}}}{t-\varepsilon_p+\varepsilon_h+\imath \eta}
    -\frac{Q^{[13]}_{\mathrm{hp}}}{t+\varepsilon_p-\varepsilon_h-\imath \eta},
    \label{eq:paulioperator13}
\end{equation}
and 
\begin{equation}
    \hat{{\cal G}}^{[14]}=
    \frac{Q^{[14]}_{\mathrm{ph}}}{u-\varepsilon_p+\varepsilon_h+\imath \eta}
    -\frac{Q^{[14]}_{\mathrm{hp}}}{u+\varepsilon_p-\varepsilon_h-\imath \eta}
    \label{eq:paulioperator14}
\end{equation}
we arrive at the following equations 
for the interaction vertex in these two channels
\begin{equation}
      \Gamma^{[13]}_{1234J}(t) = 
      V^{[13]}_{1234J}+
      \sum_{ph}
      V^{[13]}_{12phJ}\hat{{\cal G}}^{[13]}
      \Gamma^{[13]}_{ph34J}(t),
      \label{eq:first13}
\end{equation}
and 
\begin{equation}
      \Gamma^{[14]}_{1234J}(u) = 
      V^{[14]}_{1234J}-
      \sum_{ph}
      V^{[14]}_{12phJ}\hat{{\cal G}}^{[14]}
      \Gamma^{[14]}_{ph34J}(u).
      \label{eq:first14}
\end{equation}
These equations, together with Eq.\ (\ref{eq:first12}),
can then form the basis for the first iteration in a self-consistent 
scheme for renormalization corrections of the Parquet type.
The origin of the minus sign in Eq.\ (\ref{eq:first14}) follows from 
the diagram rules \cite{kstop81} and will be examplified below.
A graphical view of these equations is given in Fig.\
\ref{fig:figs1314}.
The reader should also keep in mind the two contributions to the particle
propagators of Eqs.\ (\ref{eq:paulioperator13}) and (\ref{eq:paulioperator14}).
See Ref.\ \cite{kt94} for a physical interpretation.

In this section we will omit a discussion of the self-energy corrections
which arise from these channels. This
will be relegated to the next section. In the nuclear
case, due to the strongly repulsive short-range character 
of the interaction, we will have 
to replace in actual calculations 
the bare interaction in the irreducible vertices $[13]$ and $[14]$
with the $G$-matrix discussed in the previous section. 
Doing this entails already a first step towards the Parquet set of equations,
in the sense that we are including short-range correlations from the 
$[12]$ channel. 
Here we will however limit the discussion to expressions 
in terms of the interaction $V$.
The aim is to try 
to recover from Eqs.\ (\ref{eq:first13}) 
and (\ref{eq:first14}) the familiar TDA and RPA equations and the 
so-called self-consistent coupled equations of Kirson \cite{kirson74}.
The hope is that these intermediate steps can bridge the 
gap between the familiar
TDA, RPA and $G$-matrix equations and the Parquet set of equations
in section \ref{sec:sec5}.

\subsection{Screened ph  and 2p2h interactions}

Here we study the screening of the particle-hole
and the 2p2h interactions given in Fig.\ \ref{fig:wavef1},
indicated by $V_{ph}$ and $V_{2p2h}$, respectively.
Before we list the final expression, it is however instructive
to consider the corrections to second order in the interaction $V$ to the ph
and 2p2h vertices. 
\begin{figure}[hbtp]
      \setlength{\unitlength}{1mm}
      \begin{picture}(100,40)
      \put(35,0){\epsfxsize=7cm \epsfbox{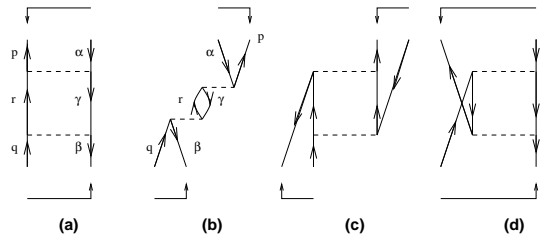}}
      \end{picture}
      \caption{Second-order perturbation theory corrections to the ph
               interaction vertex.}
      \label{fig:phvertex}
\end{figure}
In Fig.\ \ref{fig:phvertex} we display the second-order corrections to
the ph diagrams of Fig.\ \ref{fig:wavef1}. 
Diagram (a) is the core-polarization correction term to
the particle-hole
interaction, and corresponds to a contribution from the 
$[14]$ channel, as indicated by the coupling order.  
The term labeled (b) corresponds to the exchange
term of (a) and is coupled in the $[13]$-order, see also
the discussion in connection with Eqs.\ (\ref{eq:ph13})
and (\ref{eq:ph14}).
The other corrections, like (c) and (d) include
particle-particle and hole-hole intermediate states, respectively.
They are irreducible in both the $[13]$ channel and the $[14]$ channel,
and can therefore enter the irreducible vertices of these two channels in 
later iterations.  
They are however not generated by various iterations of Eqs.\
(\ref{eq:first13}) 
and (\ref{eq:first14}). In fact, if we replace $V$ by $G$ in Eqs.\
(\ref{eq:first13}) 
and (\ref{eq:first14}), diagram (c) is already accounted for by the 
$G$-matrix. It may however be included if the double-partitioned $G$-matrix
of the previous section is used. 
Let us now look at the analytical expressions 
in an angular momentum coupled basis for diagrams (a) and (b)
of Fig.\ \ref{fig:phvertex}. Here we just include the first term of
the propagators of Eqs.\ (\ref{eq:paulioperator13}) and
(\ref{eq:paulioperator14}). The second terms will give rise to the
2p2h contributions discussed below. In the following discussion we will
also assume that the interaction $V$ does not depend on the energy,
although 
it is rather easy to generalize to an energy dependent 
interaction.
Diagram (a) reads
\begin{equation}
      (a)=-\sum_{r\gamma}(-)^{j_r+j_{\gamma}-J}(-)^{2j_{\gamma}}
      V^{[14]}_{p\gamma r\alpha J}
      \frac{1}{u+\varepsilon_{\gamma}-
                \varepsilon_{r}} V^{[14]}_{r\beta q\gamma J},
       \label{eq:secordph}
\end{equation}
The factor $(-)^{2j_{\gamma}}$ stems from the opening up
and recoupling of an internal particle-hole pair \cite{kstop81}
and the phase $(-)^{j_r+j_{\gamma}-J}$ is needed in order
to rewrite the matrix elements in the coupling order of
Eq.\ (\ref{eq:13channel}).
The general structure of Eq.\ (\ref{eq:secordph}) is
just of the form $-V_{\mathrm{ph}}^{[14]}Q^{[14]}_{\mathrm{ph}}/\epsilon^{[14]}V_{\mathrm{ph}}^{[14]}$, with
$ \epsilon^{[14]}=\varepsilon_{q}+\varepsilon_{\gamma}-\varepsilon_{\beta}-
                \varepsilon_{r}=u+\varepsilon_{\gamma}-\varepsilon_{r}$
and we have defined
\begin{equation}
  u=\varepsilon_{q}-\varepsilon_{\beta}=\varepsilon_{p}-\varepsilon_{\alpha},
\end{equation}
for the on-shell energy case.
This is the equivalent of the energy variable of Eq.\ (\ref{eq:energy12}) in
the $[12]$ channel.
Diagram (b) is in turn given by
\begin{equation}
      (b)=\sum_{r\gamma}(-)^{j_r+j_{\gamma}-J}(-)^{2j_{\gamma}}
      V^{[13]}_{\gamma pr\alpha J}
      \frac{1}{\epsilon^{[13]}}V^{[13]}_{\beta rq\gamma J} ,
       \label{eq:secordphdirect}
\end{equation}
and we note that the contributions are clearly different.
The minus sign in Eq.\ (\ref{eq:secordph}) stems from the standard
diagram rules \cite{kstop81}. 
In our use of the diagram rules below, we will omit the use
of the rule for the number of external valence hole lines.
In our case then, as can also be deduced from inspection of
Fig.\ \ref{fig:phvertex}, diagram (a) has  zero closed loops and
three hole lines,
giving thereby rise to a minus sign. 
Diagram (b) has an additional closed
loop and thereby yielding the plus sign.
The energy denominator is in this case
\begin{equation} 
      \epsilon^{[13]}=t+\varepsilon_{\gamma}-\varepsilon_{r},
\end{equation}
with 
\begin{equation}
  t=\varepsilon_{q}-\varepsilon_{\beta}=\varepsilon_{p}-\varepsilon_{\alpha}.
\end{equation}
We notice, using the relations discussed
in Eqs.\ (\ref{eq:ph13})
and (\ref{eq:ph14}), that diagram (a) is simply the 
exchange diagram of (b). We need however to include
both diagrams in order to obtain an
antisymmetric equation for the particle-hole channels
which exhibits the same properties as the $[12]$ channel
shown in Eq.\ (\ref{eq:symproperties}). {\em This is actually crucial
in solving the Parquet equations. We wish namely that 
every iteration, with a given approximation to the 
vertex function $V$, preserves the antisymmetry property.}
This point cannot be emphasized enough. Let us now see what 
happens to third order in the interaction.
\begin{figure}[hbtp]
      \setlength{\unitlength}{1mm}
      \begin{picture}(100,40)
      \put(35,0){\epsfxsize=7cm \epsfbox{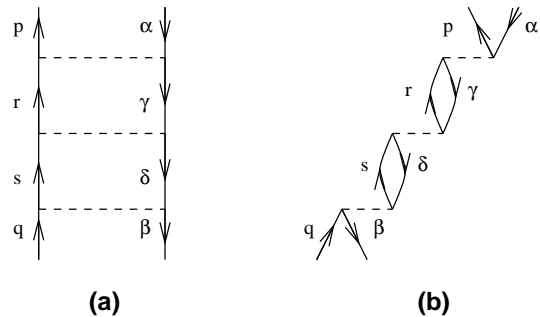}}
      \end{picture}
      \caption{Corrections beyond second order in the interaction $V$ 
               to the ph
               interaction vertex. (a) is in the $[14]$ channel and (b) is in
               $[13]$ channel.}
      \label{fig:phhigher}
\end{figure}
Third order corrections to the ph vertices (a) and (b)   
 involving only ph intermediate states are
shown in (a) and (b) of Fig.\ \ref{fig:phhigher}, respectively.
The analytical expression for the third-order contribution (a) is given by
\begin{equation}
      (a)=\sum_{rs\gamma\delta}f
      V^{[14]}_{p\gamma r\alpha J}
      \frac{1}{u+\varepsilon_{\gamma}
       -\varepsilon_{r} } V^{[14]}_{r\delta s\gamma J}
      \frac{1}{u+\varepsilon_{\delta}
       -\varepsilon_{s} } V^{[14]}_{s\beta q\delta J},
       \label{eq:thirdpha}
\end{equation}
with $f=(-)^{j_r+j_s+j_{\gamma}+j_{\delta}-2J}
      (-)^{2j_{\gamma}+2j_{\delta}}$. This
equation has the general structure
\[
            V_{\mathrm{ph}}^{[14]}
            \frac{Q_{\mathrm{ph}}^{[14]}}{\epsilon^{[14]}}
             V_{\mathrm{ph}}^{[14]}
            \frac{Q_{\mathrm{ph}}^{[14]}}{\epsilon^{[14]}}
            V_{\mathrm{ph}}^{[14]}.
\]
A similar expression applies to diagram (b), whose expression
is
\begin{equation}
      (b)=\sum_{rs\gamma\delta}f
      V^{[13]}_{\gamma pr\alpha J}
      \frac{1}{t+\varepsilon_{\gamma}
        -\varepsilon_{r} } V^{[13]}_{\gamma rs\delta J}
      \frac{1}{t+\varepsilon_{\delta}
               -\varepsilon_{s} } V^{[13]}_{\beta s q\delta J}.
       \label{eq:thirdphb}
\end{equation}
It has the general structure
\[
            V_{\mathrm{ph}}^{[13]}
            \frac{Q_{\mathrm{ph}}^{[13]}}{\epsilon^{[13]}}
             V_{\mathrm{ph}}^{[13]}
            \frac{Q_{\mathrm{ph}}^{[13]}}{\epsilon^{[13]}}
            V_{\mathrm{ph}}^{[13]}.
\]
But these expressions have the same sign! Diagram (a) counts
now 4 hole lines, and (b) counts also 4 hole lines and 2 closed
loops. However, {\em there are three interaction terms $V$}, 
and taking
the exchange term of each of these in diagram (a) leads to
the desired results, namely $(a)=-(b)$, as it should.
Thus, to third order we keep the antisymmetry property of 
$\Gamma$ in the $[13]$ and $[14]$ channels.
It is easy to see that in the $[14]$ channel we will always
have an alternating sign in front of each contribution, 
since every new order in perturbation theory brings a new hole
line and no closed loop, and thus a new minus sign.
In the $[13]$ channel we have always one new hole line and
one new closed loop for every new vertex. 
If we consider only the screening of the ph vertex, we can then set up
a perturbative expansion in terms of the ph vertex for the 
vertex functions $\Gamma^{[13]}$ and $\Gamma^{[14]}$. 
For notational economy we will skip the Pauli operators
$Q_{\mathrm{ph,hp}}^{[ij]}$ in the discussions below.
It will always be understood that the intermediate states
are two-body particle-hole states, $\left| \mathrm{ph}\right\rangle$ or
$\left| \mathrm{hp}\right\rangle$. 
Consider e.g.,
$\Gamma^{[14]}$ 
\begin{equation}
       \Gamma^{[14]}=V^{[14]}_{\mathrm{ph}}-
        V^{[14]}_{\mathrm{ph}}
        \frac{1}{\epsilon^{[14]}}
        V_{\mathrm{ph}}^{[14]}+
        V^{[14]}_{\mathrm{ph}}
        \frac{1}{\epsilon^{[14]}}
        V^{[14]}_{\mathrm{ph}}
        \frac{1}{\epsilon^{[14]}}
        V^{[14]}_{\mathrm{ph}}-+\dots,
\end{equation}
which can be summed up to yield 
\begin{equation}
  \Gamma^{[14]}=V^{[14]}_{\mathrm{ph}}-
   V^{[14]}_{\mathrm{ph}}
   \frac{1}
   {\epsilon^{[14]}-V^{[14]}_{\mathrm{ph}}}
   V^{[14]}_{\mathrm{ph}}=
   V^{[14]}_{\mathrm{ph}}-
   V^{[14]}_{\mathrm{ph}}
   \frac{1}{\epsilon^{[14]}}\Gamma^{[14]},
   \label{eq:screening1}
\end{equation}
which is the standard TDA expression for the ph term.
The corresponding expression in the $[13]$ channel 
results in
\begin{equation}
  \Gamma^{[13]}=V^{[13]}_{\mathrm{ph}}+
   V^{[13]}_{\mathrm{ph}}
   \frac{1}{\epsilon^{[13]}}\Gamma^{[13]}.
\end{equation}
The signs agree with the 
expressions of Blaizot and Ripka \cite{br86}, see chapter 15 and
Eq.\ (15.50).
The summations in both channels ensures that the final vertex
is antisymmetric and the combination of 
the latter two equations results in the familiar
TDA equations, see e.g., Ref.\ \cite{kt94} for a matrix equation
version. 
We next look at the $2p2h$ matrix element and show the
corresponding corrections to second order in perturbation
theory in Fig.\ \ref{fig:pphhvertex}.
\begin{figure}[hbtp]
      \setlength{\unitlength}{1mm}
      \begin{picture}(100,60)
      \put(35,0){\epsfxsize=8cm \epsfbox{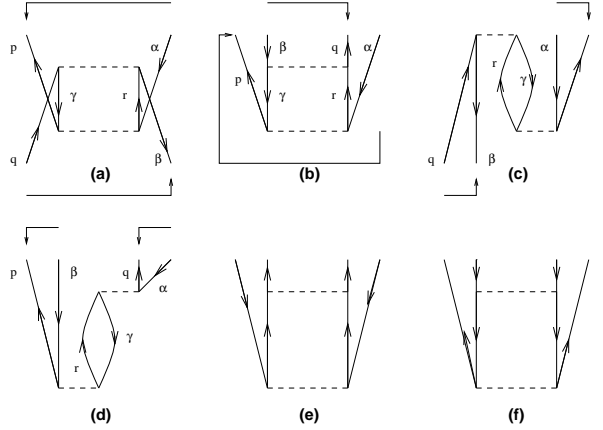}}
      \end{picture}
       \caption{Corrections to second order in $V$ of the 2p2h
               vertex.}
       \label{fig:pphhvertex}
\end{figure}
If we omit diagrams (e) and (f) which contain 2p and 2h 
intermediate states generated by the solutions
in $[12]$ channel, we have 
for diagram (a)
\begin{equation}
      (a)=-\sum_{r\gamma}(-)^{j_r+j_{\gamma}-J}
      (-)^{2j_{\gamma}}
      V^{[14]}_{\gamma\beta qr J}
      \frac{1}{-u+\varepsilon_{\gamma}-
                \varepsilon_{r}} V^{[14]}_{pr\gamma\alpha J},
       \label{eq:2p2ha}
\end{equation}
with the general structure
\begin{equation}
    -V_{\mathrm{2p2h}}^{[14]}
     \frac{1}{\epsilon^{[14]}}
     V_{\mathrm{2p2h}}^{[14]}.
\end{equation}
Note well the minus sign in front of $u$. The contribution from the 
propagator can in this case be retraced to the second
term of the propagator of Eq.\ (\ref{eq:paulioperator14}).
Diagram (a) follows the coupling order of the $[14]$ channel.
It is also easy to see that diagram (b) is given by
\begin{equation}
      (b)=-\sum_{r\gamma}(-)^{j_r+j_{\gamma}-J}
      (-)^{2j_{\gamma}}
      V^{[14]}_{\gamma q \beta r J}
      \frac{1}{-u+\varepsilon_{\gamma}-
                \varepsilon_{r}} V^{[14]}_{pr\gamma\alpha J},
       \label{eq:2p2hb}
\end{equation}
and has the structure
\begin{equation}
    -V_{\mathrm{ph}}^{[14]}
     \frac{1}{\epsilon^{[14]}}
     V_{\mathrm{2p2h}}^{[14]}.
\end{equation}
Similarly, if we now move to the $[13]$ channel we have 
the following expressions
\begin{equation}
      (c)=\sum_{r\gamma}(-)^{j_r+j_{\gamma}-J}
      (-)^{2j_{\gamma}}
      V^{[13]}_{\beta\gamma qr J}
      \frac{1}{-t+\varepsilon_{\gamma}-
                \varepsilon_{r}} V^{[13]}_{qp\gamma\alpha J},
       \label{eq:2p2hc}
\end{equation}
and 
\begin{equation}
      (d)=\sum_{r\gamma}(-)^{j_r+j_{\gamma}-J}
      (-)^{2j_{\gamma}}
      V^{[13]}_{\gamma q r\alpha J}
      \frac{1}{-t+\varepsilon_{\gamma}-
                \varepsilon_{r}} V^{[13]}_{pr\beta\gamma J},
       \label{eq:2p2hd}
\end{equation}
with the general structure
\begin{equation}
    V_{\mathrm{2p2h}}^{[13]}
     \frac{1}{\epsilon^{[13]}}
     V_{\mathrm{2p2h}}^{[13]},
\end{equation}
and
\begin{equation}
    V_{\mathrm{ph}}^{[13]}
     \frac{1}{\epsilon^{[13]}}
     V_{\mathrm{2p2h}}^{[13]},
\end{equation}
respectively. Diagram (c) is just the exchange of (a)
and includes two 2p2h vertices, while diagram (d) is the exchange
of diagram (b) and includes a ph vertex multiplied with
a 2p2h vertex. We note again that the antisymmetry is
ensured at a given order in the interaction only
if we include the corrections at the same level in both
channels.

One can then easily sum up higher-order corrections 
to the 2p2h diagrams as well in both channels.
The inclusion of the backward going particle-hole
pair in the propagators of Eqs.\ (\ref{eq:paulioperator13})
and (\ref{eq:paulioperator14}) ensures thus that we will
also sum to infinite order 2p2h corrections.
This leads ultimately to the familiar RPA equations,
see e.g., Refs.\ \cite{eo77,kt94}.  

A closer inspection of Eqs.\ (\ref{eq:2p2hb}) and
(\ref{eq:2p2hd}) shows that if we only include ph vertices,
we could resum these corrections to infinite
order for the 2p2h vertex by observing that the 
structure of such diagrams would be of the form
(e.g., in the $[14]$ channel )
\begin{equation}
       \Gamma_{\mathrm{2p2h}}^{[14]}=V^{[14]}_{\mathrm{2p2h}}-
        V^{[14]}_{\mathrm{ph}}
        \frac{1}{\epsilon^{[14]}}
        V_{\mathrm{2p2h}}^{[14]}+
        V^{[14]}_{\mathrm{ph}}
        \frac{1}{\epsilon^{[14]}}
        V^{[14]}_{\mathrm{ph}}
        \frac{1}{\epsilon^{[14]}}
        V^{[14]}_{\mathrm{2p2h}}.
             +\dots,
\end{equation}
which can be summed up to yield 
\begin{equation}
  \Gamma^{[14]}_{\mathrm{2p2h}}=V^{[14]}_{\mathrm{2p2h}}-
   V^{[14]}_{\mathrm{ph}}
   \frac{1}
   {\epsilon^{[14]}-V^{[14]}_{\mathrm{ph}}}
   V^{[14]}_{\mathrm{2p2h}},
   \label{eq:screening2}
\end{equation}
and similarly for the $[13]$ channel, but with a plus sign.
The modification discussed in Eqs.\ (\ref{eq:screening1})
and (\ref{eq:screening2})
serve to modify the propagation of a particle-hole pair
and have normally been termed for propagator
renormalizations, as can easily be seen from 
Eqs.\ (\ref{eq:screening1})
and (\ref{eq:screening2}) where the propagation of 
a free particle-hole pair is modified by the presence 
of the interaction $V$ in the energy denominator.
Other important processes which can affect e.g.,
various polarization terms are those 
represented by so-called vertex renormalizations,
a term originally introduced by Kirson and Zamick \cite{kz70}.
These authors studied the renormalizations of the 
2p1h and 2h1p vertices as well, see also Kirson \cite{kirson74}
and Ellis and Osnes \cite{eo77} for further discussions.
We will therefore end the discussion in this section
by looking at such renormalizations.

\subsection{Further renormalizations}

In the previous subsection we dealt mainly with what has 
conventionally been labelled for propagator
renormalizations. We will therefore extend the standard TDA and RPA 
scheme by looking at further ways of renormalizing   
interaction vertices. 
The approach discussed here follows 
Kirson \cite{kirson74}. Extensions were made later 
by Ellis and Goodin \cite {eg80} and
Ellis, Mavrommatis and M\"uther \cite{emm91}. 
We will limit the discussion here to the scheme of Kirson.
We start therefore with the contributions to second
order to the 2p1h vertex\footnote{The discussion here applies to the 
other interaction vertices discussed in Fig.\ \ref{fig:wavef1}.}.
\begin{figure}[hbtp]
      \setlength{\unitlength}{1mm}
      \begin{picture}(100,60)
      \put(35,0){\epsfxsize=8cm \epsfbox{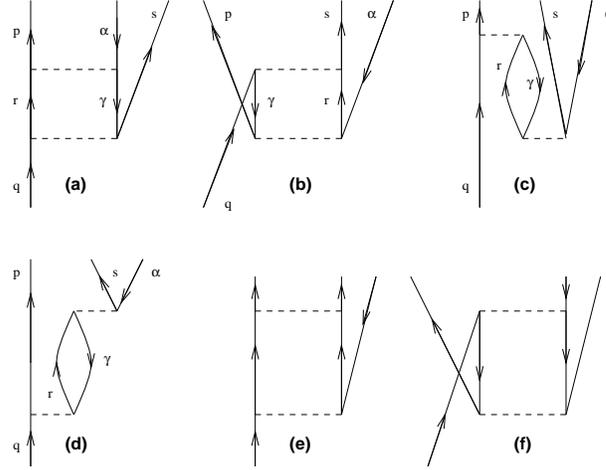}}
      \end{picture}
       \caption{The corrections to second order in $V$ of the 2p1h
               vertex.}
       \label{fig:2p1hvertex}
\end{figure}
These contributions are shown in Fig.\ \ref{fig:2p1hvertex}. 
Diagram (a) consists of a 2p1h vertex multiplied with a 
ph vertex whereas (b) stems from the multiplication
of a 2p2h vertex with a 2p1h vertex. They both contain
a particle-hole pair as an intermediate state and
follow the  coupling order of the $[14]$ channel.
The exchange diagram of (b)  is given by 
(c), while that of (a) ig diagram (d). Diagrams (e) and (f)
represent contributions from the $[12]$ channel
and are irreducible in both particle-hole channels. 
Before we sketch the 
general structure of the renormalization 
procedure of Kirson, it is instructive to consider
again the equations to second order in perturbation theory,
as the general expressions can be deduced from inspection
of these diagrams.
Diagram (a) is
\begin{equation}
      (a)=\sum_{r\gamma}(-)^{j_r+j_{\gamma}-J}
      (-)^{2j_{\gamma}}
      V^{[14]}_{p\gamma r\alpha J}
      \frac{1}{u+\varepsilon_{\gamma}-
                \varepsilon_{r}} V^{[14]}_{rsq\gamma J},
       \label{eq:2p1ha}
\end{equation}
and the plus sign stems 
from the diagram rules \cite{kstop81}, 
i.e., we
have two hole lines and no closed loop.
The propagator is that arising
from the first term in Eq.\ (\ref{eq:paulioperator14}).
The  general structure is
\begin{equation}
     V_{\mathrm{ph}}^{[14]}
     \frac{1}{\epsilon^{[14]}}
     V_{\mathrm{2p1h}}^{[14]}.
     \label{eq:2p1hseca}
\end{equation}
Diagram (b) reads
\begin{equation}
      (b)=\sum_{r\gamma}(-)^{j_r+j_{\gamma}-J}
      (-)^{2j_{\gamma}}
      V^{[14]}_{pr\gamma\alpha J}
      \frac{1}{-u+\varepsilon_{\gamma}-
                \varepsilon_r} V^{[14]}_{\gamma sqr J},
       \label{eq:2p1hb}
\end{equation}
with the following structure
\begin{equation}
     V_{\mathrm{2p1h}}^{[14]}
     \frac{1}{\epsilon^{[14]}}
     V_{\mathrm{2p2h}}^{[14]}.
     \label{eq:2p1hsecb}
\end{equation}
In this case the propagator stems from the second term
in Eq.\ (\ref{eq:paulioperator14}).
Similar equations arise for e.g.,  the 2h1p
vertices of Fig.\ \ref{fig:wavef1}. 

\begin{figure}[hbtp]
      \setlength{\unitlength}{1mm}
      \begin{picture}(100,50)
      \put(35,0){\epsfxsize=8cm \epsfbox{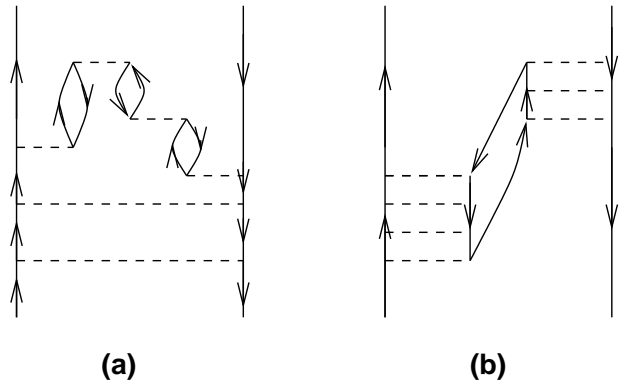}}
      \end{picture}
       \caption{Examples of diagrams which can arise from Kirson's
                self-consistent equations.}
       \label{fig:kirsoniterate}
\end{figure}

Before we write down the self-consistent equations
of Kirson \cite{kirson74}, let us assume that we
can approximate the 2p1h vertex in the $[14]$ channel 
by the first order term and diagrams (a) and (b).
This renormalized vertex, which we here label
$\tilde{V}_{2p1h}$, is then given by
\begin{equation}
     \tilde{V}_{2p1h}\approx {V}_{2p1h}^{[14]}
     +V_{\mathrm{ph}}^{[14]}
     \frac{1}{\epsilon^{[14]}}
     V_{\mathrm{2p1h}}^{[14]}+
     V_{\mathrm{2p1h}}^{[14]}
     \frac{1}{\epsilon^{[14]}}
     V_{\mathrm{2p2h}}^{[14]}.
     \label{eq:2p1hsecondorder}     
\end{equation} 
If we now allow for the screening to infinite order
of the ph vertex given by Eq.\ (\ref{eq:screening1})
and replace the 2p2h vertex in the above equation
with Eq.\ (\ref{eq:screening2}) we obtain the following
renormalization of the 2p1h vertex
\begin{equation}
     \tilde{V}_{2p1h}= {V}_{2p1h}^{[14]}
     +V_{\mathrm{ph}}^{[14]}
     \frac{1}
     {\epsilon^{[14]}-V_{\mathrm{ph}}^{[14]}}
     V_{\mathrm{2p1h}}^{[14]}+
     V_{\mathrm{2p1h}}^{[14]}
     \frac{1}
     {\epsilon^{[14]}-V_{\mathrm{ph}}^{[14]}}
     V_{\mathrm{2p2h}}^{[14]}.
     \label{eq:screening3}
\end{equation} 
A similar equation applies to the 2h1p vertex of
Fig.\ \ref{fig:wavef1} and
for the $[13]$ channel. Eqs.\ (\ref{eq:screening1}),
(\ref{eq:screening2}) and (\ref{eq:screening3}) form then
the starting point for the approach of Kirson \cite{kirson74}.
Examples of diagrams which can be obtained through the 
iterative solution of Eqs.\ (\ref{eq:screening1}),
(\ref{eq:screening2}) and (\ref{eq:screening3})
are given in Fig.\ \ref{fig:kirsoniterate}.

The question now is however how to relate 
Eqs.\ (\ref{eq:screening1}),
(\ref{eq:screening2}) and (\ref{eq:screening3}) with those
from Eqs.\ (\ref{eq:first13}) and
(\ref{eq:first14}).
This is rather trivial if we recall that the labels
$1234$ can, as was also discussed in section \ref{sec:sec2},
represent whatever single-particle states, either holes
or particles. Thus, $V_{1234}$ can represent
a 2p1h, 2h1p, 2p2h, 2p, 2h or a ph vertex.
This means that, due to the choice of propagators
in Eqs.\ (\ref{eq:paulioperator13}) and 
(\ref{eq:paulioperator14}) equations like Eq.\ 
(\ref{eq:screening3}) are already inherent in Eqs.\
(\ref{eq:first13}) and
(\ref{eq:first14}). If we e.g., approximate Eq.\ (\ref{eq:first14})
to second order in the interaction $V$ and let the 
single-particle labels $1234$ represent a 2p1h interaction
vertex, we immediately reobtain Eq.\ 
(\ref{eq:2p1hsecondorder}). If we let $1234$ represent
a 2p2h vertex, we find to second order diagrams     
(a)-(d) of Fig.\ \ref{fig:pphhvertex}.

Till now we have however refrained from discussing the contributions
from the $[12]$ channel, examples were only shown in (c) and (d) of
Fig.\ \ref{fig:phvertex}, (e) and (f) of Figs.\ \ref{fig:pphhvertex}
and \ref{fig:2p1hvertex}. 
These diagrams cannot be generated by simply iterating
the equations for the $[13]$ and $[14]$ channels, but they could enter
as contributions to the irreducible vertex 
in the $[13]$ and $[14]$ channels from the first iteration
in the $[12]$-channel. 
We see thus the emerging contour of an iterative scheme.
The crucial point is however how to perform the next iteration of say
Eqs.\ (\ref{eq:screening1}),
(\ref{eq:screening2}) and (\ref{eq:screening3}) or
Eqs.\ (\ref{eq:first13}),
(\ref{eq:first14})  and (\ref{eq:first12}) from the 
$[12]$ channel
The question is how do we include the results
from the first iteration into the next one, i.e.,
how to modify the bare vertices $V^{[13]}$ and $V^{[14]}$
in e.g., Eqs.\ (\ref{eq:first13}) and
(\ref{eq:first14}) in order to obtain
an effective interaction for the shell model.
We have also not addressed how to deal
with the solution of Dyson's equation for the one-body
Green's function. 
We mention also that Ellis and Goodin \cite{eg80} 
included pp correlations, i.e., terms from the $[12]$ channel
such as diagram (e) of Fig.\ \ref{fig:2p1hvertex},
as well when they considered the screening of the 2p1h and 2h1p vertices.
Furthermore, as already mentioned in the introduction 
the authors of Refs.\ \cite{emm91,hmm95}
extended the pp RPA to include the particle-hole (ph) RPA, though
screening of the 2p1h and 2h1p vertices was not included. In Ref.\ 
\cite{hmm95} however, a study with 
self-consistent  single-particle energies was also performed.
These works represent thus a first serious step towards the solution
of the Parquet equations, i.e., a many-body scheme which solves
self-consistently the equations in the $[12]$, $[13]$ and $[14]$
channels, with the addition of the self-consistent evaluation
of the self-energy. It ought also to be mentioned 
that one of the really
first applications for nuclear systems was performed 
in a series of papers by Dickhoff and M\"uther and co-workers
\cite{nuclearmatter}  for nuclear matter. These authors
actually performed the first iteration of the three
channels.

Such a  synthesis of the equations for the 
three channels discussed here
will be made in the next section.
 
\section{Effective interactions for finite nuclei 
         from Parquet diagrams}
\label{sec:sec5}

The equations we discussed in the two previous sections can 
be generalized
in matrix form as
\begin{equation} 
      \Gamma= \Gamma^{[ij]}+\Gamma^{[ij]}\hat{{\cal G}}^{[ij]}\Gamma,
      \label{eq:generalchannel}
\end{equation}
where obviously $[ij]$ represents a given channel,
$\hat{{\cal G}}^{[ij]}$ is the particle-particle, or hole-hole or particle-hole
propagator. 
The propagator is a product of two single-particle propagators $g$
which we do not specify any further here. They are
defined by the solution of Dyson's equation in Eq.\ (\ref{eq:dyson12}).
The irreducible vertices must appear in the solution of
the self-energy, and conversely, the self-energy must appear in all
single-particle propagators within the expressions for the three
channels $[12]$, $[13]$ and $[14]$. 

For all of our practical purposes, the irreducible vertex used in all
channels is the so-called free $G$-matrix defined in Eq.\
(\ref{eq:freeg}). We will explain more why this is our preferred choice
below.
Let us now define the contribution from the $[12]$, $[13]$ and
$[14]$ channels following Ref.\ \cite{jls82}. 
Eq.\ (\ref{eq:first12}) is then rewritten as
\begin{equation}
    L=\Gamma^{[12]}-G_F,
\end{equation}
where obviously $L$ stands for ladder and if we neglect hole-hole
terms in Eq.\ (\ref{eq:first12}) we obtain 
\begin{equation}
    L=\Delta G =G-G_F,
\end{equation}
that is the ladder corrections beyond first order in our 
irreducible vertex function $G_F$, as discussed in section \ref{sec:sec3}.
The ladder term can then be rewritten as 
\begin{equation}
    L=G_F\hat{{\cal G}}^{[12]}G_F+G_F\hat{{\cal G}}^{[12]}L.
    \label{eq:ladder}
\end{equation}
In a similar way we can define the diagrams beyond
first order in the particle-hole channel as
\begin{equation}
    R^{[13]}=\Gamma^{[13]}-G_F,
\end{equation}
and 
\begin{equation}
    R^{[14]}=\Gamma^{[14]}-G_F,
\end{equation}
where $G_F$ now is coupled in either the $[13]$ or $[14]$ way.
Rewriting $R^{[ij]}$, where $R$ refers to ring diagrams,
we obtain
\begin{equation}
    R^{[ij]}=G_F\hat{{\cal G}}^{[ij]}G_F+G_F\hat{{\cal G}}^{[ij]}R^{[ij]},
    \label{eq:ring}
\end{equation}
where $ij$ stands for either $[13]$ or $[14]$.
The equation for the vertex function $\Gamma$ in Eq.\
(\ref{eq:generalchannel}) becomes then
\begin{equation}
    \Gamma=G_F+L+R^{[13]}+R^{[14]}.
     \label{eq:gammap}
\end{equation}
{\em Note here that the vertex $\Gamma$ can be represented in the coupling
order of any of the above channels}. Our convention is that of the 
$[12]$ channel.

If Eqs.\ (\ref{eq:ladder}) and (\ref{eq:ring}) define the first iteration,
it should be fairly obvious to see that the next iteration would be
\begin{equation}
    L=\left(G_F+R^{[13]}+R^{[14]}\right)\hat{{\cal G}}^{[12]}\left(G_F+R^{[13]}+R^{[14]}\right)
      +\left(G_F+R^{[13]}+R^{[14]}\right)\hat{{\cal G}}^{[12]}L,
    \label{eq:laddernext}
\end{equation}
where the contributions $R^{[13]}+R^{[14]} $ are recoupled according to the coupling order
of the $12$-channel. These contributions are irreducible in the $[12]$ channel.
Similarly, for the rings we have
\begin{equation}
    R^{[13]}=\left(G_F+L+R^{[14]}\right)\hat{{\cal G}}^{[13]}
             \left(G_F+L+R^{[14]}\right) + 
             \left(G_F+L+R^{[14]}\right)\hat{{\cal G}}^{[13]}R^{[14]},
    \label{eq:ring13next}
\end{equation}
and
\begin{equation}
    R^{[14]}=\left(G_F+L+R^{[13]}\right)\hat{{\cal G}}^{[14]}
             \left(G_F+L+R^{[13]}\right) + 
             \left(G_F+L+R^{[13]}\right)\hat{{\cal G}}^{[14]}R^{[14]}.
    \label{eq:ring14next}
\end{equation}

Our scheme for calculating $\Gamma$ will be an iterative one based on 
Eqs.\ (\ref{eq:gammap})-(\ref{eq:ring14next}) 
and the solution of Dyson's equation
for the single-particle propagator. This set of equations will then yield
the two-body Parquet diagrams. 
Relating the above equations to the discussions of sections \ref{sec:sec3}
and \ref{sec:sec4}, it is rather easy to see 
that the $G$-matrix, TDA, RPA and Kirson's
screening scheme are contained in Eqs.\ 
(\ref{eq:gammap})-(\ref{eq:ring14next}) .

\subsection{Petit Parquet}

The aim here is to present a numerically viable approach
to the Parquet equations.
Here
we will limit ourself to just sketch 
the structure of the solution, more technical details 
will be represented elsewhere \cite{mhj99}.

The iterative scheme starts with the solution of Eq.\ (\ref{eq:first12}).
The bare vertex which is irreducible in all three channels
is the so-called free $G$-matrix $G_F$ defined in Eq.\ (\ref{eq:freeg}).
Let us use the identity from Ref.\ \cite{bbp63} and rewrite the
vertex function $\Gamma^{[12]}$ as
\begin{equation}
    \Gamma^{[12]}=G_F+G_F\left(\frac{Q^{[12]}}{s-H_0}
                               -\frac{1}{s-H_0}\right)\Gamma^{[12]},
\end{equation}
where 
\[
  G_F = V+V\frac{1}{s-H_0}G_F.
\]
We define then $Q^{[12]}=Q_{pp}+Q_{hh}$ and use
Eq.\ (\ref{eq:matrix_relation_q}) to rewrite $Q_{pp}$ and obtain
\begin{equation}
   \Gamma^{[12]}_{(0)}=G_F-G_F\left(\frac{Q_{hh}}{s-H_0}+
   \frac{1}{s-H_0}\tilde{P}\frac{1}{\tilde{P}(s-H_0)^{-1}
   \tilde{P}}\tilde{P}\frac{1}{s-H_0}\right)\Gamma^{[12]}_{(0)}.
   \label{eq:approx12channel}
\end{equation}
The subscript $(0)$ means that this is just the first iteration.
The single-particle energies are the unperturbed
harmonic oscillator energies.  

Since we will always deal with real single-particle energies,
the two-body propagators of Eqs.\ (\ref{eq:paulioperator12}),
(\ref{eq:paulioperator13}) and (\ref{eq:paulioperator14})
will have the same expressions except for the fact that
the single-particle energies get renormalized after each 
iteration. This means that 
the denominators 
in $G_F$, $\Gamma^{[12]}_{(0)}$ and subsequent iterations
and the expressions for the ring diagrams from the $[13]$ and 
$[14]$ channels, 
can be rewritten  via the simple relation
\begin{equation}
     \frac{1}{s\pm \imath\eta}=P\frac{1}{s}\mp\imath\pi\delta(s).
     \label{eq:principalvalue}
\end{equation}
With this caveat, we can in turn obtain the real and imaginary parts
of all matrices involved.

The first step in our calculations is to evaluate $G_F$.  It is 
calculated in momentum space for a series of starting
energies (typically $\sim 30$).
The principal value integral from Eq.\ (\ref{eq:principalvalue})
is solved  using Kowalski's
\cite{kowalski67} 
method. This method  ensures a numerical 
stable treatment with large numbers of
mesh points in momentum space. From the principal value integral we
can in turn define the real and imaginary part of $G_F$, see 
e.g., Ref.\ \cite{rpd89} for technical details.
The matrix $G_F$ is solved only once and transformed
to a harmonic oscillator basis in the lab from the 
relative and center of mass system, see Ref.\ \cite{hko95}
for details.

The second step is to solve Eq.\ (\ref{eq:approx12channel}),
which is now a complex equation. 
Since the number of two-hole states is rather limited,
typically  $\le 100$
for a given $J$ even for heavy nuclei like Pb, the major
problem in the matrix inversion of Eq.\ (\ref{eq:approx12channel})
resides in the dimensionality of $\tilde{P}$ discussed in section
\ref{sec:sec3}. What conditions our truncation of $\tilde{P}$ 
and the number $n_3$, is dictated by the 
Brueckner-Hartree-Fock (BHF) independence on the chosen oscillator parameter
$b$. This choice leads to the inclusion of more than ten major shells
in the computation of the $G$-matrix.
For a given $J$ value, the total number of two-body states
needed can  then be of the order $ \sim 10^4$.

{\em Our first approximation} is therefore to truncate the available
space of single-particle states to be within $\sim 10-20$ major shells.
The choice being conditioned by the BHF criteria and the possibility
to store these large matrices in the RAM of available computing
facilities.  
If we choose to use a double-partitioned model space, we need 
to sum up further diagrams with particle-particle intermediate states.

These two steps lead then to the first iteration of
the ladders, i.e.,
\begin{equation}
     L_{(0)}=\Gamma^{[12]}_{(0)}-G_F.
\end{equation} 
It contains both hole-hole and particle-particle intermediate
states and is a complex matrix. The external single-particle
legs can be particles or holes. Only unperturbed single-particle 
energies enter the definition of the two-body propagators.

The third step is to calculate the first iteration for the  
rings, namely
\begin{equation}
    R_{(0)}^{[13]}=\left(G_F+L_{(0)}\right)\hat{{\cal G}}^{[13]}
             \left(G_F+L_{(0)}\right) + 
             \left(G_F+L_{(0)}\right)\hat{{\cal G}}^{[13]}R_{(0)}^{[13]},
    \label{eq:ring13first}
\end{equation}
and
\begin{equation}
    R_{(0)}^{[14]}=\left(G_F+L_{(0)}\right)\hat{{\cal G}}^{[14]}
             \left(G_F+L_{(0)}\right) + 
             \left(G_F+L_{(0)}\right)\hat{{\cal G}}^{[14]}R_{(0)}^{[14]}.
    \label{eq:ring14first}
\end{equation}
The equations for $L$ and $R$ are all defined within a
truncated Hilbert space. They can therefore be recast into
matrix equations of finite dimensionality.
Recall also  that we need to recouple the contribution
from the $[12]$ into the relevant ones for the $[13]$ and $[14]$
channels. This is done employing Eqs.\ (\ref{eq:13channel}) and
(\ref{eq:14channel}). 
With these contributions, we can now obtain the vertex function
$\Gamma$ after the first interaction
\begin{equation}
    \Gamma_{(0)}=G_F+L_{(0)}+R_{(0)}^{[13]}+R_{(0)}^{[14]}.
\end{equation}
The fourth step is to compute the self-energy and thereby obtain
new single-particle energies. In so doing, care has to be exercised
in order to avoid double-counting problems. A thourough discussion
of this topic can be found in Ref.\ \cite{jls82}. More details
will also be presented in Ref.\ \cite{mhj99}.
Through dispersion relations \cite{rpd89} we can in turn obtain
the real part of the self-energy and our single-particle energies
will be approximated by
\begin{equation}
  \varepsilon_{\alpha}=t_{\alpha}+\mathrm{Re}\Sigma_{\alpha}.
\end{equation}
This is {\em our second approximation}.
The new single-particle wave functions of 
Eq.\ (\ref{eq:selfconstbasis}) are obtained
by diagonalizing a matrix of dimension 
$n_{\alpha}\times n_{\alpha}$, $n_{\alpha}$ the quantum
number $n$ of the single-particle state $\alpha$. 

The fifth step is to repeat steps 1-4 with the new single-particle
energies till a predetermined self-consistency is obtained. 
But now the rings have to be included in all equations, i.e.,
we solve Eqs.\ (\ref{eq:gammap})-(\ref{eq:ring14next}). 

The final vertex $\Gamma$ can then be used to define a 
new effective interaction to be applied in shell model studies,
where many more diagrams are considered than in present 
state of the art calculations, see e.g., Fig.\ 8 of Ref.\ \cite{jls82}
for a list of diagrams to sixth order entering the definition
of  the irreducible vertex $\Gamma$.

\section{Perspectives}

\label{sec:sec6}

The reader should always 
keep in mind that the many-body scheme we have focussed on 
is only one of several
possible ways of calculating effective interactions. Other methods
are also discussed in this volume.

The aim of this work has however been to show how one can 
practically implement
the Parquet equations in order to obtain effective interactions
for the nuclear shell model. The emphasis here has been to connect
these equations 
with state of the art approaches to 
effective interactions. Thus, 
how to recover the standard $G$-matrix equation and
the TDA and RPA equations from the Parquet equations has been
outlined. Applications of this many-body scheme will be presented
elsewhere \cite{mhj99}, although in depth discussions and applications
of both
the $G$-matrix, folded diagrams and other perturbative 
resummations can be found in e.g., Refs.\ \cite{hko95,eo77}.
There are also subtle technical details which deal with 
double-counting problems in the outlined iterative Parquet scheme
that we have left out, due to space limits, 
in the discussion. We take the liberty here to
refer to e.g., Refs.\ \cite{jls82,scalapino}. They will also be discussed
in Ref.\ \cite{mhj99}.

In the scheme we have sketched, there are obviously other important
many-body contributions at the two-body level
which cannot be generated by the two-body Parquet equations.
We have also omitted any discussion of three-body terms.
Such terms could be generated if
we were to solve the three-body Parquet equations, see Lande and
Smith \cite{jls82}.
An implementation of the three-body Parquet equations is in progress.
Three-body terms, as also mentioned in the introduction, are believed
to be important, it should just  suffice
to mention studies of the Triton \cite{nogga97}.
With two-body forces only one is also not able
to reproduce properly the nuclear matter saturation point
\cite{apr98}. 
In connection with three-body contributions one has to carefully
distinguish between three-body forces and effective three-body
interactions. To understand this point consider the following arguments
from studies of infinite nuclear matter and finite nuclei.
The last three years have seen quite some advances in 
many-body studies of dense infinite matter, see e.g., Refs.\ 
\cite{apr98,engvik97,baldo98} for recent surveys. 
These results can be summarized as follows. 
Firstly, new $NN$  interactions such as
the CD-Bonn potential \cite{cdbonn}, different Nijmegen interactions
\cite{nijmegen94} and the recent Argonne $V_{18}$ interaction
\cite{argv18}
all fit 
the set of scattering data of the 
Nijmegen group 
with a $\chi^2$ per datum close to 1. All these interactions,
when applied to calculations of the equation of state (EoS),
yield essentially similar equations of state up to
densities of $3-4$ nuclear matter saturation density  
for both pure neutron matter
and $\beta$-stable matter when 
the non-relativistic lowest-order Brueckner theory (LOB) is used. 
Other properties like the 
symmetry energy and proton fractions do also show
a similar quantitative agreement, see Ref.\ \cite{engvik97} for more
details.
Secondly, the inclusion of more complicated 
many-body terms at the two-body \cite{apr98}
level does not alter this picture and even the recent 
summation of three-hole line diagrams of Baldo and co-workers
\cite{baldo98} results in an EoS which is  
close to LOB when a continuous
choice is used for the single-particle energies in matter
\cite{baldo98}. The latter are examples of effective
three-body contributions. These findings 
are also in line with
recent works on the energy of pure neutron 
drops, where three-body clusters are included \cite{ndrops97},
and large-scale shell-model calculations of Sn 
isotopes including 
effective three-body interactions \cite{eho99}.
Differences do however occur 
when one introduces real three-body forces.
These are necessary
in order to reproduce the saturation properties of nuclear matter
\cite{apr98} and the binding energy of light nuclei \cite{vijay}.

The reader may then eventually ask why do we bother at all to solve
the two-body set of Parquet equations when most likely these new
effective interactions will not cure any of the problems
seen in shell model studies or nuclear matter. One needs at least to
include some realistic three-body force and thereby to solve
the three-body Parquet equations. 
The problem however with real three-body forces is that presently
we have no serious candidates which exhibit the same level
of quality and sophistication 
as the $NN$  interactions mentioned above.
The introduction of such forces lead therefore to strong model
dependencies. Moreover, to assess properly many-body terms at the two-body
level is an important, and not yet solved, problem per se.
The Parquet equations allow one also to include in a self-consistent way
several many-body terms, fulfill crossing symmetries and
satisfy certain sum rules. In addition, the Green's function formalism
can be taylored for finite temperatures. 
This is clearly of interest for studies of infinite matter at finite
temperature, such as e.g., dense matter occuring in a newly born
neutron star. 
The application of the Parquet equation in nuclear and neutron matter will also
allow for a consistent treatment of screening effects relevant for
the pairing problem. Such work is in progress.

\subsection*{Acknowledgements}
I have greatly benefitted from many discussions on Parquet theory with 
Andy Jackson. Moreover, many interactions with David Dean, Paul Ellis,
Torgeir Engeland, Tom Kuo, 
Herbert M\"uther, Artur Polls, Eivind Osnes and Andres Zuker 
have hopefully matured
some of the ideas exposed, although any eventual flaw(s) is(are) obviously
to be retraced to the author.

\end{document}